\documentstyle[11pt,aaspp4]{article} 

\slugcomment{Accepted for publication in the Astrophysical Journal}

\newcommand{\kms}{{\,\rm km\,s}^{-1}} 
 
\lefthead{Saha {\rm et~al.}}
\righthead{SN~1991T in NGC~4527}

\def\placetable#1{\vspace{0.5ex}\begin{center}EDITOR: PLACE TABLE \ref{#1} HERE.
\end{center}\vspace{0.5ex}}
\def\placefigure#1{\vspace{0.5ex}\begin{center}EDITOR: PLACE FIGURE \ref{#1}
HERE. \end{center}\vspace{0.5ex}}
\makeatletter
\renewcommand{\fps@figure}{htbp}         
\renewcommand{\fps@table}{htbp}
\makeatother

\begin{document}

\title{Cepheid Calibration of the Peak Brightness of SNe~Ia. \\ X.
       SN~1991T in NGC~4527 \\}

\author{A. Saha} 
\affil{National Optical Astronomy Observatories \\
950 North Cherry Ave., Tucson, AZ 85726} 
\author{Allan Sandage}
\affil{The Observatories of the Carnegie Institution of Washington \\
813 Santa Barbara Street, Pasadena, CA 91101} 
\author{Frank Thim, Lukas Labhardt, G. A. Tammann} 
\affil{Astronomisches Institut der Universit\"at Basel \\ 
Venusstrasse 7, CH-4102 Binningen, Switzerland} 
\author{J. Christensen}
\affil{National Optical Astronomy Observatories \\
950 North Cherry Ave., Tucson, AZ 85726}
\and
\author{ N. Panagia\altaffilmark{1}, F.D. Macchetto\altaffilmark{1}}
\affil{Space Telescope Science Institute \\ 3700 San Martin Drive,
Baltimore, MD 21218}

\altaffiltext{1}{Affiliated to the Astrophysics Division, Space
                 Sciences Department of ESA.}

\begin{abstract}

Repeated imaging observations have been made of NGC~4527 with the {\it
Hubble Space Telescope} between April and June 1999, over
an interval of 69 days.  Images were obtained on 12 epochs in the
$F555W$ band and on five epochs in the $F814W$ band.  The galaxy
hosted the type Ia supernova SN1991T, which showed relatively unusual 
behavior by having both an abnormal spectrum near light maximum, and a
slower declining light curve than the proto-typical Branch normal 
SNe~Ia.

  A total of 86 variables that are putative Cepheids have been found, with 
periods ranging from 7.4 days to over 70 days. From photometry with the 
DoPHOT program, 
the de-reddened distance modulus is determined to be $(m -M)_0 =
30.67 \pm 0.12 $ (internal uncertainty) using a subset of
the Cepheid data whose reddening and error parameters are secure. A parallel 
analysis of the Cepheids using photometry with ROMAFOT yields 
$(m -M)_0 =30.82 \pm 0.11 $. The final adopted modulus is 
$(m -M)_0 =30.74 \pm 0.12 \pm 0.12 $ ($ d=14.1 \pm 0.8 \pm 0.8 $ Mpc). 

The photometric data for SN1991T are used in combination with the  
Cepheid distance to NGC~4527 to obtain the absolute magnitude for this 
supernova of $M_{V}^{0}(max) = -19.85 \pm 0.29$. 
The relatively large uncertainty is a result of the range in estimates
of the reddening to the supernova. Thus SN1991T is seen to be only 
moderately brighter (by $\sim 0.3$ mag) than the mean for spectroscopically 
normal supernovae, although magnitude differences of up to 0.6 mag cannot be 
ruled out.  

\end{abstract}

\keywords{Cepheids --- distance scale --- galaxies: individual
(NGC~4527) --- supernovae: individual (SN~1991T)} 

\section{Introduction} 

This is the tenth paper of a series whose purpose is to obtain Cepheid 
distances to galaxies that have produced supernovae of type Ia (SNe~Ia), 
thereby calibrating their absolute magnitudes at maximum light. 

\subsection{Homogeneity and Diversity among SNe~Ia}

From the tightness of the Hubble diagram, it was recognized early on that many 
type I supernovae have nearly the same absolute magnitude at maximum, and 
that they would be useful in obtaining distances to galaxies that produced 
them. If true, supernovae could play a major role in the calibration of the 
extra-galactic distance scale (\cite{kow68}), leading to the Hubble constant 
quite directly. The only intermediate step required would ``simply'' be to 
calibrate their mean absolute magnitude at maximum if that mean was strictly
a constant, or to devise means to determine the second-parameter corrections 
if there is a range in absolute magnitude (i.e. a significant diversity). 

The evidence for the spectral homogeneity of a subclass of supernovae 
eventually called type I was progressively discovered. The evidence began with 
the exhaustive discussion of the spectra of SN~1937C in IC~4182, and SN~1937D
in NGC~1003 by \cite{min39}. His further discussion of supernovae spectra 
showed differences from the prototype spectra he had studied in 1939. As a
result, he introduced the type I and type II classification that became 
standard (\cite{min41,min64}). 

More subtle differences in the spectra, even within the type I class, began 
to be recognized in 1964. \cite{ber64}, \cite{bersu65}, and \cite{ber65} had 
noticed the lack of certain spectral features (no $\lambda \, 6150 $ \AA \, 
absorption 
and no P~Cygni profiles) in a few SNe~I, contrary to most of the class, but 
no further action was taken at that time. However, by 1985 the variations 
discovered by Bertola were shown to be general, and the SNe~Ia class was 
divided into the sub-types Ia and Ib 
(\cite{pan85,wheelev85,uokir85,bra86,har87}). 
The class was formally named by \cite{eli85}, and formally defined by
\cite{porfi87}. 

Diversity was soon again noticed even in the new Ib class. 
Harkness et~al.\ (1987) 
and \cite{harwh90} showed that the type Ib should again be divided into two 
separate groups on the basis of spectra. They introduced the new class of 
Ic. 

The stability (constancy) of the mean absolute magnitude of the classical 
SN~Ia type was also being questioned. Progressive improvements in the 
photometry of SNe since the pioneering work of \cite{baa38} and \cite{baaz38}
had shown the remarkable similarity of the shape of the light curves of 
SNe~Ia. However, by 1973 it began to be noticed that small systematic
differences 
in fact do exist in the details of the light curves, in particular, in 
their decay rates after maximum light. \cite{bar73} gave an important summary,
following earlier indications of a decay-rate absolute magnitude 
correlation by Pskovskii (1967, 1971, 1984). There were also indications that 
the expansion velocities of the SNe~Ia are not all the same and that the 
differences are correlated with absolute magnitude and decay-rate 
(Branch 1981, 1982). 

However, at the same time as spectral diversities were being found
that required division of  
the Minkowski broad class I into three groups, improvements in the photometry
and the discovery of many more SNe~Ia at relatively large redshifts 
($v > 3000 \, {\rm kms^{-1}}$) permitted major improvements in the formulation 
of the Hubble diagram (redshift vs. apparent magnitude at maximum) of SNe~Ia.

The distribution of magnitude residuals about the linear regression of 
$m$ vs. $5 \log v$ for a linear velocity-distance relation of course 
immediately gives the rms scatter of the mean absolute magnitude, 
$<\!M\!>_{\rm SNe~Ia}$. This Hubble diagram scatter became progressively smaller 
as the 
photometry improved (Sandage \& Tammann 1982, 1993, 1997), showing that 
any systematic deviation in $<\!M\!>_{\rm SNe~Ia}$ must be smaller than 
$\sim 0.4$ mag 
for Branch normal (\cite{bra93}) SNe~Ia, to which the samples that defined 
the cited Hubble diagrams had been restricted.

A history of this early work and the pros and cons of believing that 
$<\!M\!>_{\rm SNe~Ia}$ was remarkably stable is given by 
Branch \& Tammann (1992), again when the 
samples are restricted to Branch normal SNe~Ia. It was because of this 
remarkable apparent constancy of $<\!M\!>_{\rm SNe~Ia}$ for samples so 
restricted that
we also restricted the absolute magnitude 
calibrations in the first nine papers of this series 
to such SNe~Ia subtypes. 

It is now known, primarily from the extensive new precise photometry 
by the supernova group at Cerro-Tololo lead by Phillips, that when the 
restriction on the spectral normalcy is removed, an appreciable range
of $<\!M\!>_{\rm SNe~Ia}$ is present. It is also now known that a small 
range in 
$M_{\max}$ exists even among the restricted Branch normal SNe~Ia. The range
depends both on decay rate and color [\cite{tri98}, \cite{saha99} (Paper~IX),
\cite{tribra99}, \cite{phi99}, \cite{par00}] confirming the decay-rate 
part of the 
correlation suggested by Pskovskii (1967, 1971, 1984) and by \cite{bar73}. 

The early paper by \cite{phi93} began the modern discussions of the 
decay-rate absolute magnitude relation. 
Tammann \& Sandage (1995), Saha et al. (1999), Parodi et al. (2000), and 
Sandage, Tammann, \& Saha (2000)
have argued that  
the steep slope in their initial formulation was overestimated. The 
magnitude of this   
dependence is now accepted to be 
milder (\cite{ham96a}) than in their initial formulation.

A more complete review of the many discovery paths from homogeneity to 
diversity from Kowal to the present, and the amplitude of each second 
parameter as it is currently understood, or is now or has been 
controversial, is discussed by Sandage, Tammann, \& Saha (2000).

\subsection{The Abnormal Spectral Case of SN~1991T}

A new development was begun with the discovery of a different kind of spectral 
abnormality in the supernova 1991T that was discovered on April 13, 1991
in NGC~4527.
The discovery was made 15 days before maximum light that occurred on April 28.
A detailed summary of the many discoveries and discussions by many different 
groups of the spectral diversity shown by SN~1991T is given in the 
Introduction in the paper by \cite{fish99}. 

Briefly, \cite{fil92} obtained spectra from 12 days before maximum to 47 days 
after maximum 
(hereafter $-12$ to $+47$d) showing that the pre-maximum spectra ``did not 
resemble those of any other supernova [but] beginning near maximum light the 
usual SNe~Ia lines of intermediate-mass elements slowly developed, and months 
after the explosion the iron-dominated spectrum appeared almost identical 
to that of a typical SN~Ia'' (Fisher et al. 1999). The same description of the 
abnormal spectra before maximum ($-13$ to $-7$d) was given by \cite{rla92},
and by    
\cite{phi92} for before and after maximum light ($-13$ to $+66$\,d). All 
independent 
data showed that the spectrum was not Branch normal before maximum light
but resembled almost identically Branch normal prototypes after maximum light.

Was the absolute magnitude at maximum light also peculiar?

\subsection{Previous Estimates of the Absolute Magnitude at Maximum of 
SN~1991T}

A large literature has developed based on the premise that SN~1991T was very 
much brighter than the usual Branch normal SNe~Ia that all have 
$<\!M\!>_{\rm SNe~Ia} = -19.48 \pm 0.07$ in both $B$ and $V$ with an rms 
dispersion 
of only 0.2 mag (Saha et al. 1999, Table~5 and equations 10 and 11). 
A number of special (non-standard) supernova explosion models 
have been discussed, some 
using absolute magnitudes as bright as $-20.2$ (e.g. Fisher et al. 1999), 
fully 0.7 mag brighter than the present canonical Branch normal mean 
absolute magnitude (see the 1999 STScI workshop on SN and Gamma Ray Bursts 
cited as Livio, Panagia \& Sahu 2000 in the references). 

However we share the opinion of many of the authors we have cited that the 
evidence for such bright absolute magnitude rests on quite insecure grounds 
for at least some of the estimates of $M(\max)_{1991T}$ in the literature. 
Even before obtaining the Cepheid data presented in this paper, our opinion
of a few of the methods used to obtain these bright estimates has led us to 
question whether $M(\max)_{1991T}$ was any more abnormal than 
$\sim 0.3$ mag from 
$<\!M(\max)\!> = -19.5$, and, therefore, whether it deviates at all 
from the shallow 
decay-rate vs. absolute magnitude relation that we derived in Paper~IX of this
series (Saha et al. 1999, Fig.~12). 

Fisher et al. (1999, their section 5) gives a good summary of some of the 
evidence used by others to infer an abnormally bright absolute magnitude for 
SN~1991T. Most unfortunately the problem lies almost entirely with the value
of the extinction suffered by SN~1991T itself. The different estimates of the
extinction by different authors, plus a variety of assumptions on the distance 
of NGC~4527 relative either to the bulk of the main Virgo cluster spirals 
or to a proposed membership in the tight X group of \cite{dev75}, which is the 
group named 11-4 by \cite{tul88} containing 
NGC~4496A, NGC~4536, and NGC~4725, also 
complicated the early estimates. One can obtain conclusions that range
from the result that $M(\max)$ of SN1991T is normal compared with 
Branch-normal 
SNe~Ia (Phillips et al. 1992 in an early paper), to those where the absolute
magnitude of SN~1991T is at least 0.6 mag brighter than the mean of normal 
SNe~Ia (Filippenko et al. 1992), or that $M^{0}_{V}(1991T)$ is 0.75 mag 
brighter than SN~1981B in NGC~4536 (Fisher et al. 1999, their Table~1). 

Our purpose in this paper is to obtain the absorption corrected distance to
the parent galaxy 
NGC~4527, thereby circumventing all assumptions of relative distances by 
group associations to other galaxies for which Cepheid distances are 
also available. 

Although we are confident that our absorption corrected Cepheid modulus of 
NGC~4527 is systematically correct to within the external uncertainties
quoted in sections 4 and 5, the conclusion of $M(\max)$ for SN~1991T still 
depends on the uncertain assumptions we make concerning its absorption.
We discuss all the possibilities based on the various assumptions concerning 
the absorption.

\section{Observations and Photometry} 
\subsection{The Data} 

Repeated images of a field in NGC~4527 were
obtained using the WFPC2 (\cite{hol95a}) on the $HST$.  
The field is shown in Fig.~\ref{fig1}, marked 
over a ground based image of NGC~4527.
The composite image of this field taken with the WFPC2 is shown 
in Fig.~\ref{fig2}. There are 12
discrete epochs in the $F555W$ passband, and 5 epochs in the $F814W$
passband, spanning a period of 69 days. The duration of this period is
constrained by the time-window during which this target can be
observed with $HST$ without altering the field orientation. 

The epochs were spaced strategically over this period to provide
maximum leverage on detecting and finding periods of Cepheid variables
over the period range 10 to 65 days. Each epoch in each filter was
made of two sub-exposures taken back-to-back on successive orbits of
the spacecraft. This allows the removal of cosmic rays by an
anti-coincidence technique described by \cite{saha96a} (Paper V). 
The images from various epochs are in common alignment to 
within 3--4 pixels on the scale of the PC chip, which is
1--2 pixels on the scale of the other three wide-field chips. The
journal of observations is given in Table~\ref{tbl1}.

\placetable{tbl1}

\subsection{Photometry}
 
\subsubsection{The Analysis done in Tucson} 

The details of processing the images, combining the sub-exposures for
each epoch while removing cosmic rays and performing the photometry
with a variant of DoPHOT (\cite{schec93}) optimized for WFPC2 data has
been given in Paper V, and are not repeated here. The
reduction procedure for the data is identical to that 
described in Paper~V, with
the one exception of a change in the definition of the ``partial
aperture'' which is discussed in Saha et al. (1999, Paper~IX).

In keeping with the precepts in Paper~V, the measurements in any one
passband are expressed in the magnitude system defined by
\cite{hol95b} that is native to the WFPC2.  These are the $F555W$ and
$F814W$ ``ground system'' magnitudes calibrated with $HST$ ``short''
exposure frames.  The issue of the discrepancy of the photometric
zero-points for the ``long'' and ``short'' WFPC2 exposures, originally
found by \cite{ste95} is described in some detail in Paper~V. 
This phenomenon is attributed to charge transfer problems in the 
WFPC2 CCDs, which is at its worst when the background level in the image 
is near zero. With increasing background levels, the problem is less present.
Thus ``short'' exposures of standard stars/fields which have near 
zero background yield photometry that is systematically different from ``long''
exposures that have higher levels of background. 
In previous papers of the series (except for IC~4182 and NGC~5253, which were 
done with WF/PC, not WFPC2) we {\it added}
$0\fm05$ in {\it both} passbands to the \cite{hol95b} calibration
whenever the exposures are longer than several hundred seconds, and 
(consequently) the background level in electrons has been relatively high.
Improved characterization of this problem as a function of 
background level and other parameters for the 
photometric procedures used in this series of papers is now available, 
and given in \cite{saha00}. However, the last word on this topic has perhaps 
not been said, particularly concerning secular trends due to continuing 
radiation damage of the CCDs.  We opt to follow the procedure in previous 
papers of nominally adding $0\fm05$ at the final stage of the reductions, 
while (as before) presenting the light curves in the {\it uncorrected} system.
In an eventual accounting, all the results can be 
treated at par, and corrected by a common prescription. 

\subsubsection{The Analysis Done in Basel}

A parallel analysis was done independently at Basel, which is 
described fully in \S5. As part of this analysis, photometry on the NGC~4527 
images was performed independently using the ROMAFOT package. 
The discussion of the Basel procedures, and the comparison with DoPHOT based 
analysis done in Tucson is postponed to \S5.

\section{Identification and Classification of the Variable Stars} 

Armed with measured magnitudes and their reported errors at all
available epochs for each star in the object list, the method
described by \cite{saha90} was used to identify variable stars.  The
details specific to WFPC2 data have been given in various degrees of
detail in Papers~V, VI, and VIII. Again, parallel efforts were made by the 
Tucson and Basel groups, and the results for identified Cepheids 
were merged. More than 90\% of the Cepheid candidates found by DoPHOT 
were also found by ROMAFOT.

All variable stars definitely identified are marked in
Fig.~\ref{fig3}.  However, some of the identified variables cannot be
seen in Fig.~\ref{fig3} because of their extreme faintness and/or
because of the large variation in surface brightness over the field.
Hence, to complement these charts, we set out in Table~\ref{tbl2} the X
and Y pixel positions for these objects as they appear in the images
identified in the $HST$ data archive as U42G0101R and U42G0102R.

\placefigure{fig3} \placetable{tbl2}

The photometry on the \cite{hol95b} ``short exposure'' calibration
system for the final list of 66 variable stars is presented in
Table~\ref{tbl3} for each epoch and each filter.  The periods were
determined with the \cite{laf65} by using only the $F555W$ passband
data.  Aliasing is not a serious problem for periods between 10 and
65 days because the observing strategy incorporated an optimum
timing scheme as before in this series. 

\placetable{tbl3}

The resulting light curves in the $F555W$ passband, together with
periods and mean magnitudes [determined by integrating the light
curves, converted to intensities, and then converting the average
back to magnitudes, and called the ``phase-weighted intensity average'' in
\cite{saha90}], are shown in Fig.~\ref{fig4}, plotted in the order of
descending period. 

\placefigure{fig4}

Most of the variable stars identified in this manner can be immediately 
recognized by their periods and light curves as Cepheids. In fact, 
none of the 66 objects are {\it a priori} inconsistent with being Cepheids, 
though there is a range in the quality of the light curves. The objects
C3-V5 and C3-V7 are borderline detections, and their amplitudes, if real, 
are extremely small. C3-V13 and C4-C23 also have very low amplitudes, but 
are more convincing as variables. C1-V1 and C2-V6 have periods larger than 
the 69 day observing baseline, and so their period determinations cannot be 
very reliable. The best estimate for the periods of these stars 
is made from the existing data from a 
subjective assessment of the implied light curve based on an assumed period. 

The available data for the variables in $F814W$ were folded with the
ephemerides derived above using the $F555W$ data.  The results are
plotted in Fig.~\ref{fig5}.  Note, however, that
four of the variables discovered from the $F555W$ photometry were not
found by our procedure in the $F814W$ images 
(C1-V12, C2-V4, C4-V7 and C4-V11). 
This is because either they are
intrinsically faint or else appear faint due to 
high extinction and  may not register clearly on the $F814W$ frames which
do not reach as faint a limiting magnitude as those in $F555W$.  Since 
photometry of such objects was obviously impossible in $F814W$, they
are dropped from Fig.~\ref{fig5} and also from further
analysis. 

\placefigure{fig5}

The mean magnitudes in $F814W$ (integrated as intensities over the
cycle) were obtained from the procedure of \cite{lab97} whereby each
$F814W$ magnitude at a randomly sampled phase is converted to a mean value
$\langle{F814W}\rangle$ using amplitude and phase information from
the more complete $F555W$ light curves.  Note that each available
observation of $F814W$ can be used independently to derive a mean
magnitude.  Hence, the scatter of the individual values about the
adopted mean $F814W$ value is an {\it external} measure of the
uncertainty in determining $\langle{F814W}\rangle$.  It is this
external measure of the uncertainty that is retained and propagated in
the later calculations. There is one instance (C1-V9) where there is only 
one available $F814W$ measurement, so only the error estimated by DoPHOT
for that observation can be propagated: in this instance the error estimate
is very high anyway, and the object plays essentially no role in the final 
distance estimate.

The prescription given in Paper~V for assigning the light-curve
quality index $QI$ (that ranges from 0 to 6) was used.  In this scheme,
two points ($0-1$) are given for the quality of the $F555W$ light curves, two
points ($1-2$) for the evenness in phase coverage of the five $F814W$
observation epochs, and three points ($1-3$) for the amplitude and phase
coherence of the $F814W$ observations compared with the $F555W$ light
curve.  Hence, a quality index of 6 indicates the best possible light
curve quality. A quality index of 2 or less indicates near fatal flaws
such as apparent phase incoherence in the two passbands.  This is
generally the indication that object confusion by crowding and/or
contamination by background is likely.

Table~\ref{tbl4} lists the characteristics of all 66 objects whose
light curves in $F555W$ are consistent with those of Cepheids.
For 62 of these, an $F814W$ measurement exists for at least one epoch.  The
$F555W$ and $F814W$ instrumental magnitudes of Table~\ref{tbl3} have
been converted to the Johnson $V$ and Cousins (Cape) $I$ standard
photometric system by the color equations used in previous papers of
this series, as set out in equations (2) and (3) of Paper~V, based on
the transformations of \cite{hol95b}.

\placetable{tbl4}

The magnitude scatter $\sigma_{\langle{V}\rangle}$ in Table~\ref{tbl4}
is based on the estimated measuring errors in the photometry of the
individual epochs.  In the case of $\sigma_{\langle{I}\rangle}$, the
{\it external} uncertainty from the \cite{lab97} procedure 
is listed, since it is more robust and realistic.  The quality
index discussed above is also listed. Other columns of
Table~\ref{tbl4} are explained later in the text.
 
\section{The Period-Luminosity Relation and the Distance Modulus} 
\subsection{The P-L Diagrams in $V$ and $I$} 

As in the previous papers of this series we adopt the P-L relation in
$V$ from Madore \& Freedman (1991) as 

\begin{equation}
 M_{V} ~~=~~ -2.76 ~\log P - 1.40~, 
\end{equation} 
whose companion relation in $I$ is 
\begin{equation}
 M_{I} ~~=~~ -3.06 ~\log P - 1.81~.  
\end{equation} 
The zero-point of equations (1) and (2) is based on an assumed LMC
modulus of 18.50.

The P-L relations in $V$ and $I$ for the 62 Cepheids in
Table~\ref{tbl4} are shown in Fig.~\ref{fig6}.  The filled circles
show objects with periods between 20 and 65 days that have a quality
index of 3 or higher.  
The filled circles show smaller scatter, and  delineate the 
P-L relation.  The solid lines show the canonical slopes 
of the P-L relations in $V$ and $I$ with the vertical offset 
for apparent distance moduli 
$\mu_{V} = 31.05$ and $\mu_{I} = 30.83$ respectively. 
These values were chosen  
to be in visual conformity with the points shown as filled circles
(they are not intended as a formal derivation of distance). 
The expected spread in each of the pass-bands due to the 
finite width of the instability strip (\cite{sata68}) is indicated by 
the flanking dashed lines.
The observed scatter of the data outside these envelope lines
are due to the combination of (1) measuring and systematic errors due
to background and contamination, (2) the random error of photon
statistics, (3) the large effects of the variable extinction
evident from the dust lanes seen in the images, and (4) objects mis-identified
as Cepheids.

The period cut-off of 20 days for the filled circles is chosen to prevent 
a bias at the faint end because Cepheids at short periods that are at the 
faint end of the intrinsic scatter about the mean P-L relation and fall below 
the detection limit in brightness do not populate the P-L relation, which  
adversely affects the fitting of the P-L relation. A period cut is unbiased, 
but the value of the shortest included period depends on the magnitude limit 
of the data. For this reason it is chosen 
independently from one galaxy to the next: nearby galaxies can accomodate 
a shorter period cut, while ones that are farther require a cut at longer 
periods for the same magnitude detection limit.

\placefigure{fig6}

\subsection{A Preliminary Analysis of the P-L Relation} 

For a first estimate, using $A_{V}/A_{I} = 1.7$
(\cite{schef82}) along with the very preliminary 
apparent moduli in $V$ and $I$ of 
31.05 and 30.83 respectively as estimated 
above yields a dereddened modulus $\mu_{0} \approx 30.50$, (all values 
quoted in sections 4.2 through 4.4 should have an 
additional 0.05 mag added for the `long vs. short' correction mentioned above
to be placed at par with the final values obtained in previous papers of this 
series). 
To explore the presence of differential extinction and 
to treat the data accordingly, we use the tools developed in Paper~V and used
again in Papers~VII (Sandage et al. 1996), VIII (Saha et al. 1997) and 
IX (Saha et al. 1999).

For each Cepheid we calculate the apparent distance moduli separately
in $V$ and in $I$ from the P-L relations of equations (1) and (2) and
the observed $\langle{V}\rangle$ and $\langle{I}\rangle$ magnitudes
from Table~\ref{tbl4}. These apparent distance moduli, called $U_{V}$
and $U_{I}$ in columns (7) and (8) of Table~\ref{tbl4}, are calculated
by
\begin{equation}
 U_{V} ~~=~~ 2.76 ~\log P + 1.40 + \langle{V}\rangle~, 
\end{equation}
and 
\begin{equation}
 U_{I} ~~=~~ 3.06 \log P + 1.81 +\langle{I}\rangle~.  
\end{equation} 
They are the same as equations (6) and (7) of Paper~V.

If the differences between the $V$ and $I$ moduli are due solely to
reddening, and if the dependence of the reddening curve on wavelength
is the normal standard dependence as in the Galaxy, then the true
modulus $U_T$ is given by 
\begin{equation}
 U_{T} ~~=~~ U_{V} - R'_{V} \cdot (U_{V} - U_{I})~, 
\end{equation} 
where $R'$ is the ratio of total to selective absorption, $A_{V}/E(V-I)$.
This is equation (8) of Paper~V. However, equation (5) is valid only
if the difference between $U_{V}$ and $U_{I}$ is due to extinction,
not to correlated measuring errors, in which case the value of $R$
would {\it not} be given by the normal extinction curve where
$A_{V}/A_{I} = 1.7$ and the ratio of absorption to reddening is
$R'_{V} = A_{V}/E(V-I) = 2.43$ (\cite{schef82}). 
It will be shown that such coupled errors clearly exist, and  
the derivation of the true distance modulus is considerably more complicated 
than would be the case in the absence of the correlated systematic
measuring errors.

The values of $U_{T}$ are listed in column 9 of Table~\ref{tbl4}.
These would be the true moduli, as corrected for normal extinction,
assuming that there are no systematic measuring errors.  The total rms
uncertainty for each $U_{T}$ value is listed in column 10.  This
uncertainty includes contributions from the estimated random measuring
errors in the mean $V$ and $I$ magnitudes, (in columns 4 and
6), as propagated through the de-reddening procedure, as well as the
uncertainty associated with the intrinsic width of the P-L relation
(i.e. a given Cepheid may not be on the mean ridge-line of the P-L
relation) as well as a ten percent uncertainty in the estimated period.
 The de-reddening procedure amplifies the measuring
errors. Therefore many Cepheids are needed to beat down these large
errors (notice the very large values in column 10) in any final value
of the modulus.  The values shown in column 10 of Table~4 were calculated 
using equation 18 of \cite{saha00}, and also correspond to $\sigma_{tot}$
as defined in Paper~V.  

It is important to mention that two regrettable chains of of error have 
propagated through our series of papers that concern the calculation of the 
uncertainties in $U_{T}$. The first is that 
equation (11) in Paper~V for the calculation of $\sigma_{width}$ 
was incorrectly given, and its correction given in 
Paper~IX was also in error: its use would be tantamount to assuming that 
the intrinsic width of the P-L relation is uncorrelated in the two 
passbands. The program that calculates these values 
was examined, and found to be coded correctly, according to equation (18) 
of \cite{saha00}. However, an error in coding the uncertainty due to 
the uncertainty in period estimates resulted in 
{\it overestimates} of $\sigma_{tot}$ in our previous papers 
that can be corrected as follows:
\begin{equation}
 \sigma_{tot}^2 (corrected)  = \sigma_{tot}^2 (old) - 0.34^2  
\end{equation} 
The values of $\sigma_{tot}$ have been used in Papers V, VIII and IX 
for calculating weighted means of $U_{T}$, using $1/\sigma_{tot}^2$
as weights. The effect of our prior miscalculations was to `flatten'
the weights, an effect that is, however, benign: it makes the 
weighted averages more like unweighted ones, although the objects with 
egregious measurement errors were still successfully down-weighted. The final 
distance moduli using weighted averages are unaffected at the $0.03$ mag level,
and unweighted averages are completely unchanged.

From the data in Table~\ref{tbl4}, the unweighted mean dereddened modulus 
$\mu_{0} = <U_{0}>$ for  all Cepheids 
with periods from 20 to 65 days with $QI \geq 3$ is $ 30.38 \pm 0.10 $ mag, 
and the weighted (by $1/\sigma_{tot}^2)$ value is $ 30.43 \pm 0.09 $ mag. 
Closer examination reveals that the Cepheids in Chip~4
systematically give lower distance moduli: the weighted average $\mu_{0}$ is 
$30.12 \pm 0.13$ for the 12 Chip~4 Cepheids alone (same period and $QI$ cut), 
but $ 30.68 \pm 0.12 $ for the 15 Cepheids in the remaining 3 Chips.
The corresponding unweighted averages are $30.08 \pm 0.12$ and 
$30.62 \pm 0.12$ respectively. 
The average (unweighted) {\it apparent} moduli $\mu_{V}$ for 
these two samples are 
$ 31.07 \pm 0.07 $ and $ 31.20 \pm 0.08 $ respectively. The  
corresponding (unweighted) values for 
$\mu_{I}$ are $30.65 \pm 0.06$ and $30.97 \pm 0.08$ for the apparent modulus.
The large discrepancy in the $I$ moduli suggests a possible problem 
with the Chip~4 photometry in $I$. This is investigated further in 
\S4.3. Including Cepheids with $QI=2$ typically drops the dereddened 
modulus by 0.1 mag, while raising the cut to higher $QI$ makes no 
noticeable difference. 

We note again that the derived $U_{T}$ values depend on the
assumption that the differences between $U_{V}$ and $U_{I}$ are due to
reddening alone, in the absence of appreciable systematic and
correlated measuring errors, or when the errors for $U_{V} - U_{I}$ {\it
are distributed symmetrically}.  
If correlated and/or asymmetrical errors in $V$ and
$I$ dominate over differential reddening,
thereby producing a ratio of the $V$-to-$I$ errors that is different from 
2.43, the $U_{T}$ derived via equation (5) {\it will be systematically 
in error}.

In particular, several Cepheids which were discovered in $V$ are too
faint in $I$ to be measured, as already mentioned.  This introduces a
selection effect that biases against Cepheids with bluer colors.  The
effect is most pronounced at short periods where the {\it intrinsic}
colors are bluest.  This effect gives an asymmetrical distribution of
errors in $U_{V} - U_{I}$ in the sense that it makes the de-reddened
modulus {\it too small}. This is likely to be more acute for the case of 
Chip~4, where the Cepheids are mostly in regions of 
extremely high surface brightness
and crowding.  

In Paper~V we devised a method to test for the presence of
differential extinction or for the fact that the scatter about the P-L
relation is due predominantly to measuring errors, or a combination of
both.  The method, shown in Fig.~11 of Paper~V for NGC~4536 and
explained in the Appendix there, has been used also in Papers VI, VIII 
and IX.  The
method is to plot the difference in the $V$ and $I$ moduli for any
given Cepheid against the $V$ modulus.  If
there is a systematic trend of the data along a line of slope
$dU_{V}/d(U_{V} - U_{I}) = 2.43$, then equation (5) applies and there
is clearly differential reddening.  If, on the other hand, there is a
general scatter with no trend, that scatter is dominated by measuring
errors. While in the latter case bona-fide differential extinction can
be hidden by measuring errors, trying to correct for putative
reddening will result in interpreting any asymmetry in the error
distribution as specious extinction.  In two of the three previous
cases, NGC~4536 (Paper~V), and NGC~4496A (Saha et al. 1996b, Paper~VI) 
there is no trend
along a differential reddening line.  In the third case of NGC~4639
(Paper~VIII), there is a slight trend but also large scatter showing
that the spread of points appears to be due to a mixture of
measurement errors as well as from differential extinction.

The diagnostic diagram just described is shown for the NGC~4527 data
from Table~\ref{tbl4} in the top panel of Fig.~\ref{fig7}. 
The filled circles show Cepheids with periods between 20 and 65
days that have $QI \geq 3$.
The solid line indicates the reddening vector for the P-L ridge
line, if the true (de-reddened) distance modulus is 30.50 which was
the initial estimate made in this section.  The dashed lines
show the bounds due to the intrinsic dispersion of the P-L relation as
explained in Paper~V.  The slope of the lines is $A_{V}/E(V-I) = 2.43$
as explained above.

\placefigure{fig7}

The scatter in Fig.~\ref{fig7} (top) is disappointing. In general the points 
do not lie within any reddening track: translating the reddening band in this 
figure would not materially reduce the number of points that would spill 
out of it. There may be some differential reddening, but the visible 
trend is that for the `good' Cepheids (filled circles) the $U_{V}$ values do 
not show any trend with $U_{V}-U_{I}$. This is a strong pointer that the 
range of $U_{V}-U_{I}$ is not due to extinction, but due to measurement
errors. The relative constancy of the $U_{V}$ values in Fig.~\ref{fig7}(top) 
suggests that the {\it differential} extinction is smaller than can be 
detected, given the errors in the data.  The range of values in $U_{V}-U_{I}$
must then be due to I-band photometry errors. This prediction can easily be 
checked by making the $I$-band analog, shown in the lower panel 
of Fig.~\ref{fig7}, where $U_{I}$ rather than $U_{V}$ is shown, and the slope 
and width of the locus of where the reddened Cepheids should lie is 
appropriately changed. The points shown with 
concentric outer circles are for Cepheids in Chip 4. 
The `good' Cepheids follow a slope 
$dU_{I}/d(U_{V}-U_{I}) \approx -1$, which confirms the prediction that the 
range in $U_{V}-U_{I}$ is due to unaccounted for errors in the $I$-band 
photometry (see section 3.2 in \cite{saha00}). Note also that most of 
the objects that are overly bright in $I$
are to be found on Chip 4, which is the most crowded part of the galaxy. 

An alternate perspective is to plot just $U_{V}$ vs. $U_{I}$, which is 
shown in Fig.~\ref{fig8}. The reddening vector is shown by the solid line, 
whereas the points seem to stretch along a slope of unity (dashed line).
In Paper~IX we discussed how confusion errors that are correlated in 
the two passbands can produce arbitrary slopes. The results of simulation
experiments performed using an identical photometry procedure
on data very similar to the one at hand (Saha et al. 2000) show that the 
slope of the correlated errors is about unity. The extreme points, which 
are expected to be outliers anyway, seem to follow the results of the 
simulations, lending credence to the hypothesis that they are the result 
of confusion errors. It is possible that the remaining points are 
partially afflicted with confusion noise, in which case 
the simple use of equation (5) 
will result in systematic errors. Additional examination of the data 
by different methods is essential, and is done in \S\S4.3 and 4.4.

\placefigure{fig8}

This is not to deny the presence of differential reddening, but 
an acknowledgement that the measurement errors in $I$ dominate the 
color spread.  We proceed by making an additional restriction of the 
data by color. Note that such a restriction used in conjunction with 
equation (5) does not introduce a procedural bias in the distance modulus.

\subsection{Inspection of the Color-Magnitude Diagram}

The observed color-magnitude diagrams (CMDs) obtained from 
DoPHOT photometry on 
the deep (combining all epochs) images is shown individually in 
Fig.~\ref{fig9} for each of the regions in the four different chips. 
Two things are noteworthy:

\begin{enumerate}

\item 
The general appearance of the CMDs from chips 1, 2 and 3 is not unlike the CMD 
shown in Fig~14 of Paper V for NGC 4536, with the difference that the 
blue edge, which is quite sharply defined for NGC~4536, shows some smearing
for the NGC~4527 field, and is placed 0.2 mag or more redder than for 
NGC~4536. This only confirms the general level 
of reddening that we have already surmised from the Cepheids, and expected, 
given the appearance of NGC~4527 with its pronounced dust lanes.

\item
It is apparent that the CMD obtained from chip~4 is different from the other
chips: there are a disproportionate number 
of stars that are more luminous than in the other chips. It would be nearly
absurd to think this a real feature, and as we show below, it indicates
a problem with the photometry at these high levels of surface brightness, 
crowding, and dust structure. 

\end{enumerate}

\placefigure{fig9}

The CMD anomaly for Chip~4 with respect to the remaining chips is seen also 
in Fig.~\ref{fig10}, where the same CMDs are plotted, but with $F814W$ on the 
ordinate. In Fig.~\ref{fig11}, the CMD for the chip~4 region alone is shown, 
but it is broken down by quadrant sub-regions. The anomaly is clearly visible 
for the upper-left and lower-left quadrants, but not noticeable in the 
two right hand quadrants. It is immediately clear from Fig.~\ref{fig2} and 
Fig.~\ref{fig3}, that the anomaly appears in the very bright areas where 
crowding is most severe, and where dust lanes cut up the background 
continuity severely. 

\placefigure{fig10}

\placefigure{fig11}

This anomaly in the photometry is not entirely unexpected. It has been 
pointed out by \cite{Moch99} and \cite{Staudal99} that photometric blending
in overcrowded fields may lead to a systematic underestimate of distances. 
However, it has also been 
the subject of thorough empirical investigation by   
Saha et al. (2000) and \cite{ferr00}.  The latter authors, who have run 
simulation experiments with artificial star frames, find no 
empirical evidence for such
large systematic drifts in photometry with crowding in excess of 
a few hundredths of a magnitude (which can lead to distance underestimates 
by $~0.1$ mag). These studies were done before data were available for 
NGC~3627 and now NGC~4527, which are by far the data-sets most 
afflicted by high surface brightness, crowding, 
and dust-lane structure. Of these 
NGC~3627 is the closer galaxy, and crowding effects are less likely to have 
been explicitly noticed. It appears that DoPHOT does not produce reliable
photometry in the overly crowded sub-regions of Chip~4 for NGC~4527. 

Recall that earlier in this paper, the scatter in the $I$ photometry for the 
Cepheids was surmised to be worse than for $V$, while the anomaly in the 
CMD appears equally bad for $V$ and $I$. The selection process for Cepheids, 
which includes not only a consideration for the light amplitude, but also a
{\it visual} inspection of each candidate Cepheid on the $V$ (but not $I$) 
image will result in rejecting objects that are likely to be mismeasured 
in $V$. However, the same objects may be contaminated seriously in $I$, 
and the equivalent checks for object rejection in $I$ were not in place. 
This may be the explanation of the empirical conclusion made earlier that 
the color-spread is dominated not by differential extinction alone, but by 
error excursions in the $I$ photometry. 

Consider now the sample of Cepheids with $20^{d} < P < 65^{d}$ and $QI \geq 3$
and including only those objects from Chip 4 that lie on the right hand 
half, away from the crowded regions ($x > 400$). This sample of 19 Cepheids
includes 4 from Chip~4, and yields a true modulus 
$(m-M)_{0} = 30.55 \pm 0.10$ using an
unweighted average, and $(m-M)_{0} = 30.57 \pm 0.11$ when weighted by the 
inverse variance. The apparent (unweighted) distance modulus in $V$ for this 
sample of Cepheids is $(m-M)_{V} = 31.20 \pm 0.08 $, and in $I$ 
we obtain $(m-M)_{I} = 30.93 \pm 0.07$. This implies an average 
characteristic reddening of $E(V-I) = 0.27$. 

\subsection{The Distance Modulus By Restricting The Data By Color} 
 
We examine the Period - Color (P-C) relation to identify Cepheids in 
our sample that lie close to the intrinsic $(<V_{0}> - <I_{0}>)$ period 
relation for unreddened Cepheids. In Fig.~\ref{fig12}, the 
P-C relation for the Cepheids in NGC~4527 are shown. Cepheids with  
$20^{d} < P < 65^{d}$ and $QI \geq 3$ are shown as filled circles. The ridge 
line for unreddened Cepheids in the Galaxy, LMC and SMC as summarized in 
\cite{sanetal99} from data by \cite{dea78}, \cite{cal85}, \cite{fern90}, is 
shown by the dashed line. The P-C relation implied by equations 1 and 2
is shown by the continuous line.  It is obvious from this figure, that hardly 
any of the Cepheids observed in NGC~4527 are unreddened.
The overall trend in the data has a steeper slope with period than 
the intrinsic P-C relation -- since the longest period objects are probably
the most strongly reddened, or because they lie in the super-crowded regions 
of Chip~4.  Note that selecting by 
color does {\it not} introduce an {\it a priori} bias in the distance 
determination (although one must never use such a sample to estimate 
reddening).

\placefigure{fig12}

When looking at 
Cepheids in a disk, we expect to see ones that are reddened very little because
they are on the near side of the dust sheet in the disk, and also those 
that are reddened substantially since they are seen through the disk and 
possibly through dust lanes. 
The rationale for a color-cut is that while bona-fide extinction should not 
affect the dereddened distance modulus irrespective of how the data are 
cut in color, photometric errors and stars mis-classified as Cepheids, which 
do affect the modulus, can be identified and rejected. A plot of $U_{T}$ 
vs. apparent color (not shown) for the data at hand easily demonstrates
that the derived distance is a strong function of the measured color, showing
again, that the data are dominated by measurement errors. 

A way to proceed is to try various cuts by color, and choose the one(s) that 
produce the P-L relations with the least scatter in both passbands. 
Following such 
a path, we were able to identify two cuts that result in P-L relations without 
significant outliers:
\begin{enumerate}
\item
A cut containing 
Cepheids bluer than $(V-I) = 1.2$ but also with $(V-I) \geq 0.7$ to remove 
blue outliers. Objects within the cut appear to follow the slope
of the fiducial P-C relation, whereas objects that appear 
redder follow an anomalously steeper slope. In any case, these are the 
Cepheids that appear to have minimal reddening. 
Further restricting the data to include only Cepheids 
for which $20^{d} < P < 65^{d}$ and $QI \geq 3$, to avoid the selection 
effect at the shortest periods (Sandage 1988) we obtain a weighted 
average from 16 objects: $ \mu_{0} = 30.69 \pm 0.11 $, and an unweighted 
average $ \mu_{0} = 30.69 \pm 0.09 $. The apparent modulus in $V$ from 
the same sample of Cepheids is $\mu_{V} = 31.08 \pm 0.05$, which is 
expected to be
biased towards brighter than average due to the color selection. 
\item
A better cut is one that runs parallel to the fiducial P-C relation, 
as shown in the figure by the dotted lines in Fig.~\ref{fig12}, 
drawn 0.2 mag below and 0.3 mag above the mean relation for unreddened 
Cepheids. The 13 Cepheids included yield $\mu_{0} = 30.64 \pm 0.12$ 
(weighted) and $\mu_{0} = 30.66 \pm 0.07$ (unweighted). The $V$ modulus 
for this sample 
yields $\mu_{V} = 31.05 \pm 0.05$, which again 
is expected to be biased towards brighter than average, since we have selected 
the least reddened Cepheids.
\end{enumerate}
Moving these cuts redward by even 0.2 mag begins to admit outliers into the 
resulting P-L relations. 

\subsection{The adopted Distance Modulus}

We have detailed two independent ways of selecting the Cepheids with 
more reliable photometry, given that there is an obvious difficulty with 
the magnitudes in the very crowded regions. One method selected objects by 
spatial position, away from the very crowded and high surface brightness 
regions which are also crossed by dust lanes.  The other uses color as a
diagnostic, on the premise that while rejecting an otherwise well measured 
but very reddened Cepheid does not introduce a systematic bias, objects 
that have extreme colors often appear so due to noise, and not because 
they are in fact reddened.  These independent approaches yield very similar 
samples of Cepheids, and the de-reddened modulus ranges from 30.55 to 
30.68, with characteristic internal rms uncertainties of order 0.12 mag, 
which is comparable to the case to case scatter. 
 
Recall that one must add 0.05 mag to bring the values at par with the 
``long vs short'' corrections applied in previous papers of this series: since 
this applies equally to both passbands, it applies directly to the 
de-reddened modulus also. Including this adjustment, it is 
reasonable to adopt the central value of the range, and carry forward 
the case to case scatter which is also the internal 
rms scatter for each case as a conservative error estimate:
\begin{equation}
(m-M)_{0} = \mu_{0} = 30.67 \pm 0.12 \pm 0.12.
\end{equation}

From the best case for the V-band apparent modulus, using the unweighted 
averages of the samples with period and quality cuts only (i.e. no color cut),
and propagating the uncertainties from the internal scatter:
\begin{equation}
(m-M)_{V} = \mu_{V} = 31.26 \pm 0.08.
\end{equation}

From a similar consideration of the $V$ and $I$ band apparent moduli, 
(but disregarding the case of the color cut which biases the colors), 
we get a characteristic average reddening for the Cepheids:
\begin{equation}
E(V-I)_{\rm Cepheids} = \mu_{V} - \mu_{I} = 0.26 \pm 0.05, 
\end{equation}
where an additional systematic calibration uncertainties of $~0.05$ mag 
should be added in quadrature. 

Fig.~\ref{fig12plus} shows the P-L relations in $V$ and $I$  again, 
but this time with the lines showing the above adopted moduli, and with 
filled circles showing the sample that have $QI \geq 3$, $20 \leq P \leq 65$, 
and where Cepheids in Chip 4 are included only if they have positions with 
$X \geq 400$. 

\placefigure{fig12plus}

\section{ROMAFOT Photometry}

\subsection{Motivation}
The uncertainty in the photometry of Cepheids from data like the one at hand
is dominated by both effects of crowding and object confusion, as well as from
the difficulty of measuring reliable aperture magnitudes for at least a 
few stars in the field of view of each CCD chip. Independent determination 
of aperture magnitudes for a given chip using only the Cepheid data 
(typically no suitable bright isolated but unsaturated stars exist in these
fields) results in aperture corrections with scatter of order 0.2 mag. 
Fortunately, the WFPC2 PSFs are stable, and the aperture corrections can be 
`imported' by different means. The method by which our DoPHOT based 
procedure does this was described in Paper~V, with an augmentation described
in Paper~IX. Nevertheless, the possibility of systematic errors in the 
method cannot be ruled out, particularly as a result of possible long-term 
drifts in the optics and instrument. 
Different crowded field programs, with their different approaches to 
handling object confusion, will produce somewhat different results. 
Intercomparing such results produces a more realistic handle on the true 
systematics. This was the underlying reason for repeating the data reduction 
in Section~4 using the ROMAFOT package, which was developed for crowded fields.
Because the ROMAFOT path provides an independent evaluation of the
photometry including the aperture correction, we averaged the results of
both paths rather than to use the ROMAFOT results as a rough confirmatory of
the DoPHOT results. 

Note that due to differences in approach between the two methods, including
different sets of Cepheids and quality indices, the 
post-photometry analysis cannot always be done identically. The methods 
best suited to the strengths and limitations of ROMAFOT are used: the final 
comparison however should reflect realistic systematic errors, albeit to the 
extent possible by comparing only two paths.

\subsection{Photometric reductions with ROMAFOT}
All observations were preprocessed by the standard pipeline as described by
\cite{hol95b}. The back-to-back exposures were then co-added by an
anticoincidence technique in order to remove cosmic-ray hits (\cite{saha96a}). 

The co-added exposures were aligned to within the nearest whole pixel
before the exposures were co-added again to create a deep frame in
each color. All 12 epochs in $F555W$ and all 5 epochs in $F814W$ were
used to create the deep images in order to get a better signal to
noise ratio. The PSFs for each epoch and filter and for the deep frames
were determined empirically from the co-added frames, mean PSFs 
were established with the observed data set, not with external data.    
This underlying philosophy implies the use of stars which cover 
the whole chip if possible in order to take into account for the 
spatially varying PSF of WFPC2. 
Only relatively bright and isolated objects were used to create the 
different PSFs.
For this purpose ROMAFOT was run in a mode in which the shape parameter of
the Moffat function was allowed to vary.  Empirical tests have shown that a
Moffat function fits the shape of the PSF better than a Gauss
function. ROMAFOT does not discriminate between stars, remaining
cosmic-rays and galaxies, therefore badly fitted objects were deleted from 
this sample after examination of the observed and the fitted shape of each
object. 

   The appropriate PSF was fitted to the deep frame, separately in $V$
and $I$. Objects brighter than a certain limit were selected and
subtracted from the deep frame. These objects are used as a master
list for every single epoch. The positions at different epochs are
slightly different to those of the deep frame. The corresponding
offsets have been calculated using a matching algorithm and are added
for every epoch.

   The objects were then fitted on the individual frames with the
corresponding PSF and subsequently subtracted from them. The
procedure was repeated on the residual frames several times in an
iterative process cutting at fainter and fainter limits. The results
are files giving instrumental magnitudes in $V$ and $I$ for
the deep frames and the individual frames, and each containing more
than 10\,000 stars for each chip and filter.

\subsection{Identification of the variable stars}

The identification of the variable stars is again based on the method
described by \cite{saha90}.  The quantities $\Theta$ and the standard
deviation $\sigma$ of the instrumental magnitudes over all epochs
in $F555W$ were used to identify the variable candidates. 
We have used here $\Theta_{\rm min}$, i.e. the lowest smoothed value,
obtained by varying the period between 1 and 90 days. Several hundred
stars were suspected to be variable on the basis of ROMAFOT
photometry. 
The light curves of these possible candidates with reasonable values
of $\Theta$ and $\sigma$ were individually inspected by eye, they were
scrutinized and 64 Cepheid candidates were retained that were also
selected by DoPHOT. 
They are listed in Table~\ref{tbl5}. 
Column~1 gives the designation
of the probable Cepheid, columns~2\,-\,4 give the
DoPHOT mean magnitudes in $V$ and $I$ and the quality class, and
columns~5\,-\,7 give the ROMAFOT mean magnitudes (as derived below) as
well as the quality index ($QI$) of the Cepheids. The $QI$s were
estimated similarly as in Section~3, but independently and based on 
ROMAFOT alone.

\placetable{tbl5}

37 Cepheids were judged to be useful from the ROMAFOT photometry (QI $\geq$
3),  26 of which are among the Cepheids which have been used for the
derivation of the distance modulus in Section 5.6.

\subsection{Transformation of instrumental to standard magnitudes}

The phase-weighted mean $<\!F555W\!>$ and $<\!F814W\!>$
magnitudes have been 
converted to Johnson $V$ and Cousins $I$ using the equations given by
\cite{hol95b}. 
In particular the ROMAFOT instrumental magnitudes were
transformed in a recursive manner. Instead of the unknown standard colors,   
first the flight colors and then the ground colors were transformed.

The main problem is to tie the fitted magnitudes to the instrumental
magnitudes using an aperture of $0\farcs{5}$ radius, the so-called
aperture correction (AC). This is because the transformations given by
\cite{hol95b} 
refer to an circular aperture of $0\farcs{5}$ radius.  
The AC is the correction from the model PSF to the circular aperture,
and its value strongly depends on the chosen sky values. 
The value of the AC should be evaluated in principle from the NGC~4527
frames for each chip, filter and epoch separately. 

Due to the lack of bright isolated stars with constant background on
NGC 4527 frames, the values for the aperture correction have also been
calculated with the globular clusters Pal4 [$HST$ Archive data, Proposal 5672, 
PI: J. Hesser] and G319
in M\,31 [$HST$ Archive data, Proposal 6671, PI: R.M. Rich]. 
The latter was observed in 1999 and lies
closest to the date of the present observations.  
All observations were `long' observations having exposure times of
more than 1000 seconds. Because of crowding, blends and varying 
background values the proper calculation of the AC is much more difficult
than to establish a PSF, furthermore external archive data were useful
in order to calculate systematic errors.  
Both globular clusters have many isolated,
bright, but not saturated stars with a flat sky. In first
approximation the growth curves of stars in the galaxy frames and the
globular cluster frames are expected to be the same, but jitter and
focus changes and aging effects of the CCD do affect the inner core of
the PSF.

   Since the determination of the sky is crucial, the local sky was
allowed to vary until the growth curves were flat. 
Only stars with flat growth curves were considered, stars with 
rising or falling growth curves up to $3\arcsec$ radius were rejected. 
No AC dependence with position on the chip, position within one pixel,
or the level of background were found. 
However there is a weak tendency, yet only on the galaxy frames, that
stars with broader FWHM yield higher ACs. The AC determined for the
galaxy frames depends therefore somewhat on the selection of the
reference stars. We have selected 20$-$30 reference stars per frame with
moderately small scatter.

 To first order, the changes in the intrinsic PSF are radial. If using a 
constant PSF were to produce systematic position dependent errors in the 
photometry, they should show up as differences between stars near the 
center of a chip versus those on the outside areas of the chip. Since 
ROMAFOT used a constant PSF, such a trend should show up as a systematic 
position dependent variation of the AC. 
An examination of the typical rms star-by-star scatter for the AC in 
(say) the Pal~4 data is $0.17$ mag. This value does not change by more 
than $0.01$ whether one 
considers the innermost $400 \times 400$ pixel region, or the whole chip.
This is an indication that systematic differences in the AC 
between inner and outer regions is no more than 0.08 mag. Given this state 
of affairs, it is not possible to evaluate any reliable AC dependence on 
position from these data. Nevertheless, since 
the mean computed AC's reflect the 
average over the entire chip, unless the Cepheids are 
distributed extremely non-uniformly, this concern can increase the 
scatter in the Cepheid to Cepheid relative moduli, but should not change 
the mean results from a sample of Cepheids.

   The value of the AC from NGC~4527 is slighty larger than from the
globular clusters. The choice of the AC has therefore some effect on
the distance modulus which is discussed in more detail in 
Section~5.6. 

Finally the magnitudes were corrected to the `long exposure' calibration
scale by adding $0.^m05$ mags in $V$ and $I$ as in \S4.5.

The two masterlists in $V$ and $I$ were joined to form a set of about
15\,000 stars with $V < 28\fm0$ for which $V$ {\em and\/} $I$ are
known. Their color-magnitude diagrams  $V-I$ versus $V$ and $I$ are
shown chip by chip in Fig.~\ref{fig13} and Fig.~\ref{fig14},
respectively. Cepheids are indicated as filled circles. 

\placefigure{fig13}

\placefigure{fig14}

\subsection{ROMAFOT versus DoPHOT photometry}
For about 10\,000 stars $V$ and $I$ magnitudes are available from
DoPHOT as well as from ROMAFOT photometry. Their magnitude differences
(all magnitudes in this discussion, and in Table~5 and later,  are on 
the `long' scale)
$\Delta V$ and $\Delta I$ in the sense ROMAFOT $-$ DoPHOT are plotted
versus $V$(ROMAFOT) and $I$(ROMAFOT), respectively, in
Fig.~\ref{fig15}. The very strong systematic differences for
stars with $V \ga 26^{\rm m}$ and $I \ga 25^{\rm m}$ are known also
from other comparisons of different photometry programs (\cite{gib00}). 
They are fully explained by mismatches within the matching radius of
$0\farcs5$ between objects identified on the $V$ and $I$ frames. As
expected the difficult Chip~4 gives the largest scatter. The zeropoint
offsets as well as the magnitude scatter of the brighter stars are
tabulated in Table~\ref{tbl6}.

\placefigure{fig15}

\placetable{tbl6}

\placetable{tbl6a}

   For the comparison, ROMAFOT magnitudes have been used which are
based on the mean AC of NGC~4527, Pal\,4, and G\,319. The mean
differences in Table~\ref{tbl6} and Table~\ref{tbl6a}, in the sense
ROMAPHOT $-$ DoPHOT, would be $0\fm18$ more negative in $V$ and
$0\fm14$ in $I$ if only NGC~4527 had been used for the AC.

Comparison of all 64 $F555W$ and 57 $F814W$ Cepheid candidates in common 
with differences less than 1 magnitude in
Table~\ref{tbl5} is shown in Fig.~\ref{fig16}. 
Some magnitude equation is apparent, the brighter Cepheids
agreeing better.
The mean values for 
the differences are $\Delta\!<\!\!V\!\!>$ 
and $\Delta\!<\!\!I\!\!>$ of $0.07 \pm 0.02$ and $0.12 \pm 0.03$,
respectively.  

\placefigure{fig16}

\subsection{The Period\,-\,Luminosity Relation and the Distance Modulus}
64 Cepheid candidates out of 66 noted by DoPHOT could also be detected
by ROMAFOT and three could not be recovered in $F814W$ . The
ROMAFOT candidates, for which both colors could be measured, were
assigned quality classes $QI$ similarly as in Section~3, but independently and
based only on ROMAFOT data. 37 Cepheids in Table~\ref{tbl5} for which
both colors could be measured have $QI \geq 3$.
The diagnostic diagram for the Cepheids from ROMAFOT photometry 
is shown in Fig~\ref{fig17} and 
their P\,-\,L
diagrams in $V$ and $I$ in Fig~\ref{fig18}.

\placefigure{fig17}
\placefigure{fig18}

   As before, the diagnostic diagram shows little differential reddening
along the reddening line, except C1-V11 and C3-V12 which seem to have
more than average reddening. C2-V6 is very faint for its long period
and is probably a W~Vir~star (Pop.\,II Cepheid). 
These three variables
as well as the outlying C4-V6 are not considered for the distance
determination. 

Again a period cut from 20 to 65 days has been applied to these 33 good 
Cepheids in order to avoid a possible bias. These 26 Cepheids with ROMAFOT 
photometry (filled circles) are shown in Fig,~\ref{fig18}.
Since the distance moduli depend somewhat on the aperture correction (AC)
adopted, different solutions are given in Table~\ref{tab:ROMAFOT_distances},
i.e.\ with the AC from NGC\,4527 proper, or from the globular clusters Pal4,
G319, NGC\,4527, Pal4 and G319, or from the mean AC of 
NGC\, 4527, Pal4 and G319 (cf. Section~5.4). Also shown is a
solution for only the Cepheids in chip~1-3 and those in chip~4.
The result from applying a color cut exactly as for case 2 in \S 4.4 is also 
shown for comparison.

The star-by-star 
fitting errors reported by ROMAFOT appear to underestimate (at least in these 
overcrowded fields) errors in the fitted 
magnitude when compared with the scatter in the values obtained 
for the same stars on repeated images. Without reliable star-by-star error 
estimates from the fitting errors, no meaningful weighting scheme can be 
applied. 
All averages calculated from the ROMAFOT data are unweighted. 
The quoted uncertainties in $\mu_{V}$ and $\mu_{I}$ reflect the random error
in measurement for each category considered. The uncertainty in
$\mu_{0}$ was calculated with the $rms$ scatter of the 
dereddened distance moduli of the individual Cepheids.
The variation from case to case as a function of adopted aperture correction 
and chip grouping is a rough indication of the systematic errors: estimates 
of 0.13 for $\mu_{V}$ and 0.08 for $\mu_{I}$ (in the mean) are indicated. 
The systematic error in $\mu_{0}$ appears to be small as gleaned by the 
case to case scatter, since the changes in the aperture corrections for 
$V$ and $I$ appear to change from case to case in a way that attenuates 
the variation in $\mu_{0}$. If we assume this to be chance coincidence, and 
that systematic errors in 
measuring the aperture corrections in $V$ and $I$ 
are uncorrelated, the systematic error in $\mu_{0}$ ought to be the one from 
propagating the systematic errors from $\mu_{V}$ and $\mu_{I}$, thus 
giving $\pm 0.27 $ mag per case. But such an estimate disregards the 
correlated color variation due to differential extinction, and is likely 
to be an overestimate. 

The best
compromise solution is apparently the one with the AC taken from
NGC\,4527, Pal4 and G319. The moduli in $V$ and $I$ lead --
with $A_{V}=2.43\,E(V-I)$ -- to the true moduli in the last column of
Table ~\ref{tab:ROMAFOT_distances}. The indicated errors of the true
moduli are dominated by the error in $E(V-I)$. This compromise solution 
is estimated to have a systematic error of approximately $\pm 0.07$ mag 
in $\mu_{0}$.

\placetable{tab:ROMAFOT_distances}

  We have presented the preferred distance estimates 
from DoPHOT (\S 4.5) and ROMAFOT (above)  values of distance modulus for 
NGC~4527. The various cases discussed for both sets of reduction/analysis 
are presented in Table~\ref{tab:ROMAFOT_distances}. For the ROMAFOT 
reductions we calculate (for the preferred aperture correction) the 
results for the cuts made for the DoPHOT data analysis, which are 
also given in Table~\ref{tab:ROMAFOT_distances}. 
The slightly larger moduli from
ROMAFOT are consistent with the fact that this photometry package in
combination with the adopted AC yields generally magnitudes roughly $0\fm1$ 
fainter than DoPHOT. Work to address this zeropoint offset is in progress, 
along with more complete simulations of ROMAFOT's performance. 

 An inspection of Table~\ref{tab:ROMAFOT_distances} shows that the preferred
estimates of the two methods, as well as estimates from cuts that eliminate 
suspect data from {\it both} methods differ by 0.11 to 0.21 mag, in the sense
that ROMAFOT gives the larger moduli. Averaging the unweighted preferred 
estimates from each approach gives $\mu_{0} = 30.74$, which is well within the
uncertainty estimates of each method (even within the {\it weighted} 
uncertainty for the DoPHOT estimates). For an overall uncertainty we could 
statistically combine the individual uncertainties from the two methods, 
but we prefer to be conservative and carry over the value common to each of 
the two methods, thus obtaining:
\begin{equation}
\label{eq:dist_mean}
 (m-M)_{0}= 30.74 \pm 0.12 \pm 0.12,
\end{equation}
which corresponds to $14.1 \pm 0.8 \pm 0.8$ Mpc. 
The systematic error is estimated from the scatter in derived dereddened moduli
for the various `Method/Selection' cases for both DoPHOT and ROMAFOT based 
reductions listed in Table~\ref{tab:ROMAFOT_distances}.
Note that the distance moduli from the acceptable (i.e. no obvious
problems identified) subsamples listed in Table~\ref{tab:ROMAFOT_distances}
range from 30.60 to 30.93, with favored values between 30.60 and 30.88. 
These numbers demonstrate the degree to which the distance determination 
is robust.

The best estimates from DoPHOT and the best compromise solution from 
ROMAFOT for $\mu_{V}$ are in remarkable
agreement, even though the samples are formally different. We average them
and retain the larger of the two uncertainties to reflect the real dispersion 
among the Cepheids. 
The adopted values for $\mu_{I}$ for the two methods 
differ by $0.09$ mag. Again we average the moduli and keep the larger 
uncertainty. The final numbers are shown in the final row of Table 8.
The systematic variations in $\mu_{V}$ and $\mu_I{}$ as seen from case to case 
within DoPHOT and ROMAFOT are not inconsistent with these adopted 
uncertainties. In particular, note that for the selected samples without a 
color-cut, $\mu_{V}$ values range from 31.12 to 31.27 from DoPHOT, and 
from 31.09 to 31.42 from ROMAFOT, since a direct $V$ band comparison 
to SN~1991T is made later \S6.1. 

The distance H\"oflich \& Khoklov (1996) derived for SN\,1991T 
from modelling its light curve and spectrum is $(m-M)_{0} = 30.4\pm0.3$.
\cite{sparks99} estimated an upper limit of 15 Mpc ($(m-M)_{0} = 30.9$)
for  
SN\,1991T, by modeling the polarized emission from its light echo. 

\section{The Peak Brightness of SN\,1991T}

\subsection{The Observed Brightness at Maximum}

The most comprehensive photometry for SN\,1991T, which was discovered
12 days before maximum light, is presented in \cite{lira98}. The
maximum brightness in $B$ and $V$ is seen to be:
\begin{equation}
  \label{eq:10}
  B_{\rm max}=11.70\pm0.02, \quad V_{\rm max}=11.51\pm0.02, \quad
  \hbox{and hence}\quad B_{\rm max}-V_{\rm max}=0.19\pm0.03.
\end{equation}
Combining the $V$ maximum with the adopted modulus $(m-M)_V = 31.27$ in
Table~\ref{tab:ROMAFOT_distances} gives

\begin{equation}
  \label{eq:11}
  M_{V}^{0}(\max)=-19.76\pm0.12.
\end{equation}
This value holds only under the assumption that SN\,1991T suffers the
same absorption as the Cepheids in NGC\,4527. The uncertainty reflects 
the realistic error in the {\it average} $\mu_{V}$ for a Cepheid: the
random error estimates for $(m-M)_{V}$ given in 
Table~\ref{tab:ROMAFOT_distances} are convolved with the statistical errors
gleaned from the different acceptable cases shown in the same table.  
An individual object can be quite different, depending on where it is located 
with respect to the dust, and so it is quite likely 
that the absorption of the SN is different. Therefore we
attempt to estimate its reddening independently in the following sub-section.

\subsection{Estimating reddening and extinction}
The Galactic absorption to NGC\,4527 is given by \cite{schle98} to
be $A_{B}=0.09$ and $A_{V}=0.07$. The total absorption must be
larger and is estimated in the following ways.

   A first approximation is obtained by assuming that SN\,1991T has
the same reddening as the Cepheids on average, i.e.\ $E(V-I)=0.22$
(mean of DoPHOT and ROMAFOT), and hence $E(B-V)=0.18$.

   If it is assumed that SN\,1991T has the same intrinsic color as
blue, Branch-normal SNe\,Ia, i.e.\ $(B_{\max}-V_{\max})=-0.01\pm0.01$ with an
intrinsic scatter of $\sigma_{B-V}=0.05$ (\cite{par00}),  one obtains
$E(B-V)=0.20\pm0.06$ with equation~(\ref{eq:10}). (An estimate of
$E(V-I)$ is not attempted, because standard SNe\,Ia have a wider
color scatter of $\sigma_{V-I}=0.08$).

   If, instead, it is argued that the spectroscopically peculiar
SN\,1991T cannot be compared with the colors of standard SNe\,Ia, it
can be compared with its spectroscopic twin SN\,1995ac
(\cite{gar96}). The latter has -- after correction for Galactic
reddening following \cite{schle98} --
$(B_{\max}-V_{\max})=-0.06$. This yields for SN\,1991T
$E(B-V)=0.25\pm0.06$ (estimated error). 

Phillips et~al.\ (1999) 
have suggested that all SNe\,Ia evolve similarly in
color at late phases in their light curves, i.e. 50 days past
maximum. They argue that since this is based on the nebular phase of
SNe, it is expected to be more robust than the color at
maximum. Accordingly, they obtain a total reddening of SN\,1991T
corresponding to $E(B-V)=0.16\pm0.05$. This value may have to be
reduced to $E(B-V)=0.12\pm0.05$ because their adopted mean color of
all SNe\,Ia is 0.04 bluer on average than found by Parodi et~al. (2000).

   From the strong Ca and Na interstellar absorption lines in the
spectrum of SN\,1991T at the radial velocity of NGC\,4527,
Ruiz-Lapuente et al. (1992) and \cite{fil92} have concluded
$E(B-V)=0.34$
and $E(B-V)=0.13-0.23$, respectively.

   The various reddening estimates are consistent with an adopted
value of $E(B-V)=0.21\pm0.08$ (i.e. the mean and standard deviation of the 
various estimates listed above) , which was also favored by
\cite{fish99}, who adopted $E(B-V) = 0.2 \pm 0.1$.
This value is not significantly different from the mean
reddening of the Cepheids in NGC\,4527. With $R_{B}=4.1$, $R_{V}=3.1$,
and $R_{I}=1.9$ one obtains then a total absorption of $A_{B}=0.82\pm0.33$,
$A_{V}=0.62\pm0.25$, and $A_{I}=0.38\pm0.15$, leading to the true
magnitudes of SN\,1991T of 
\begin{equation}
  \label{eq:12}
  B_{\max}^{0}=10.88\pm0.33, \quad V_{\max}^{0}=10.89\pm0.25,
  \;\hbox{and}\quad I_{\max}^{0}=11.29\pm0.16.
\end{equation}
Together with the true modulus of NGC\,4527 from Table~8, this
yields absolute magnitudes of
\begin{equation}
  \label{eq:13}
  M_{B}^{0}({\max})=-19.86\pm0.36, \quad M_{V}^{0}({\max})=-19.85\pm0.29,
  \;\hbox{and}\quad M_{I}^{0}({\max})=-19.45\pm0.21.
\end{equation}
When these magnitudes are compared with the mean absolute magnitude of
eight blue standard SNe\,Ia, whose luminosities are calibrated through
Cepheids (\cite{saha99}); one finds an overluminosity of SN\,1991T of 
\begin{equation}
  \label{eq:14}
  \Delta{M}_{B}=0.37\pm0.37, \quad \Delta{M}_{V}=0.37\pm0.30,
  \hbox{and}\quad \Delta{M}_{I}=0.20\pm0.22.
\end{equation}
These differences are further reduced by $\sim0.05$ if one compares
SN\,1991T with standard SNe\,Ia of equally slow lightcurve decline of,
$\Delta m_{15}=0.95$ [cf. \cite{par00}, eq.~(3)\,-\,(5)].
Yet the comparison remains inconclusive because of the large errors of
the absolute magnitudes of SN\,1991T which are dominated by the
uncertainty in $E(B-V)$. Within the errors it is possible that $M_{B}({\max})$
and $M_{V}({\max})$ are up to $\sim0\fm6$ brighter than for normal
SNe\,Ia.  However, a high value would cause difficulties with $M_{I}({\max})$. 
From equation~(\ref{eq:13}) SN\,1991T is, with $(V-I)=-0.40$, the bluest
SN\,Ia known, and any increase of $E(B-V)$ above the adopted value
can make it only bluer.

   It might be argued that SN\,1991T, as a peculiar SN\,Ia, may be as
blue in $(V-I)$ as it wants. However, its spectroscopic twin
SN\,1995ac (see below) has $(V-I)=-0.24\pm0.08$, after correction for
Galactic absorption, which is close to the mean value of normal
SNe\,Ia of $<\!V-I\!>\; = -0.29\pm0.02$ (\cite{par00}).
 
   Of course it is more telling to compare the bolometric luminosities
instead of the absolute magnitudes in different pass bands. Standard
SNe\,Ia with $<\!M_V\!>=-19.48\pm0.07$ (\cite{saha99}) and a
bolometric correction of $0.1\pm0.1$ (\cite{hof95}; \cite{maz95};
\cite{nug95}; \cite{bra97}) have $M_{\rm bol}=-19.38\pm0.12$.    
SN\,1991T had a larger fraction of its energy in the near ultraviolet
near maximum and its bolometric correction is therefore $-0.1\pm0.1$
(Fisher et~al.\ 1999), and hence with equation~(14) $M_{\rm
  bol}(SN\,1991T)=-19.95\pm0.31$, which is brighter by $0.57\pm0.33$
than standard SNe\,Ia. It lies not in the scope of this paper to
decide whether the (poorly determined) excess luminosity of SNe\,Ia
can be fuelled by a Chandrasekhar mass or not (cf. Fisher et~al.\
1999).

\section{Conclusions}


\subsection{The surprisingly small distance of NGC\,4527}
NGC~4527 lies in de Vaucouleurs' X-cloud, which is part of
the W-Cloud (Binggeli, Popescu \& Tammann 1993), southwest
of the Virgo cluster and outside the isophlets and X-ray contour of the
cluster. 

The W-Cloud is known to have a considerable depth (Federspiel, Tamman \&
Sandage 1998). It is therefore not too surprising that the distance of
NGC~4527 of $(m -M)_{0} = 30.74 \pm 0.12 \pm 0.12$  
($14.1 \pm 0.8 \pm 0.8$ Mpc) is even somewhat
smaller by 0.3 - 0.4 mag than the Cepheid distances of the two neighboring
galaxies NGC~4536 and NGC~4496A (Saha et al. 1996a, 1996b). The three
galaxies have - with observed velocities near $1750 \kms$ (Sandage \&
Tammann 1987) - large peculiar velocities. They lie apparently on the near
side of the Virgo cluster, probably infalling for the first time (Tully \&
Shaya 1984).

\subsection{The SN\,1991T class of SNe\,Ia}
The consequence of the small Cepheid distance of NGC\,4527 is that the
suspected overluminosity (Fisher et~al.\ 1999 and references therein) 
of SN\,1991T becomes marginal. The excess luminosity in $M_B$, $M_V$,
and $M_I$ depends entirely on the adopted internal extinction
$E(B-V)$. However, because of its negative bolometric correction
SN\,1991T seems to have a higher bolometric luminosity by $\Delta
M_{\rm bol}=0.57\pm0.33$ than standard SNe\,Ia, a difference still
barely at the 2 sigma level.

   A few other SNe\,Ia are known which have similar spectra, i.e.\
SN\,1997cw (\cite{gar97}), SN\,1998ab (\cite{gar98}), SN\,1998es
(\cite{jha98}), and SN\,1999cw (\cite{riz99}), as well as SN\,1995bd
and SN\,1995ac (\cite{gar96}), and SN\,1997br (\cite{qiao97}) for which
additional photometry is available. Some photometric parameters of the
three latter are compared with SN\,1991T in Table~\ref{tbl:SNe}.

\placetable{tbl:SNe}

   It is a curious fact that SN\,1991T received much attention because
of its seemingly high luminosity which is now marginal at best in
the light of the new Cepheid distance, while its spectroscopic twin
SN\,1995ac, which was discovered later, appears to be considerably
overluminous indeed. It should be noted that the overluminosity of
SN\,1995ac is independent of any adopted value of $H_0$ because it
lies $0\fm5$ above the Hubble line of standard SNe\,Ia and its
recession velocity of $14\,700\kms$ cannot significantly be altered by
peculiar motions. 

   SN\,1995bd does not shed additional light on the discrepancy
between SN\,1991T and SN\,1995ac. It is strongly absorbed by Galactic
dust, i.e.\ $A_{V}=1.65$ according to \cite{schle98}, but remains
somewhat underluminous and very red in $(B-V)$ after a corresponding
correction. If it is assumed that it suffers additional internal
absorption and that its color $(B-V)$ is that of standard SNe\,Ia and
similar to its two counterparts, it becomes extremely
overluminous. The latter conclusion holds for any reasonable value of
$H_0$. Even adopting an extreme value of  $H_0=80$ makes it have 
$M_{B}(\max)=-20.0$. For $H_{0}=60$, the absolute magnitude becomes
$M_{B}(\max)=-20.62$ using again the large absorption correction.

   Unfortunately the case of SN\,1997br is even more ambiguous.
\cite{li99} find this object more reddened than SN\,1991T by $\Delta
E(B-V)=0.21$, which with the color excess adopted for the latter gives
$E(B-V)=0.41$ for SN\,1997br in good agreement with $E(B-V)=0.39$ from
the Na absorption line (Li et~al.\ 1999). With $E(B-V)\approx0.41$
SN\,1997br becomes clearly overluminous. However, we have adopted the
individual recession velocity of ESO 576\,-\,G40, host of SN\,1997br,
namely $v_{0}=2070\pm20\kms$ (Li et~al. 1999) which becomes
$v_{220}=2199\kms$ after correction for Virgocentric
infall. \cite{li99} have instead adopted the mean velocity of
$1583\kms$ of the NGC\,5084/87 group, of which ESO 576\,-\,G40 may be
a member. In addition they assume only $E(B-V)=0.35$. With these
different precepts SN\,1997br becomes $0\fm90$ fainter than listed in
Table~\ref{tbl:SNe} and the SN takes roughly the luminosities and
colors of standard SNe\,Ia.

   In spite of all uncertainties it seems unavoidable that
SN\,1991T-like objects form a very heterogeneous class. Even if one
treats the extinction as a free parameter they differ significantly
either in luminosity or in color or both. Some may require more than
the Chandrasekhar mass, like SN\,1995ac, others not. It is remarkable
that three out of four SNe\,Ia in Table~\ref{tbl:SNe} have an internal
extinction of probably $E(B-V)>0.20$ which is quite unusual among
known standard SNe\,Ia.

\acknowledgments{Acknowledgments: }
We thank the many individuals at STScI who worked hard behind the scenes to 
make these observations possible. We also thank an insightful referee whose 
comments have improved this paper. 
A.S. and A.S. acknowledge support from NASA through grant GO-5427.02-93A
from the Space Telescope Science Institute, which is operated by the 
Association of Universities for Research in Astronomy. 
L.L. and G.A.T. thank the Swiss National Science Foundation for continued 
support.

\clearpage


%
%
\clearpage
%


\figcaption[]{Ground-based image of NGC~4527. 
The field covered by the WFPC2 of the $HST$ is superposed.  
North is up, and East is to the left.
The position of SN~1991T is marked.
\label{fig1}}

\figcaption[]{Color image of the $HST$ field made by stacking
several frames in both the $V$ and $I$ passbands and combining.
\label{fig2}}

\figcaption[]{Identifications for all the variable stars found.
The numbers are the same as in Tables~ \ref{tbl2}, \ref{tbl3}, \ref{tbl4} and 
\ref{tbl5}.  
Each of the four WFPC2 chips are shown separately.
\label{fig3}}

\figcaption[]{Light curves, plotted in order of period, in the $F555W$
band.    
\label{fig4}}

\figcaption[]{Same as Fig.~\ref{fig4} but for the $F814W$
passband, adopting the periods and the phasing used in
Fig.~\ref{fig4}.
\label{fig5}}
 
\figcaption[]{The apparent P-L relations in $V$ and $I$ showing
all the Cepheid data in Table~\ref{tbl4}.  The solid circles are
Cepheids with $ 65^{d} > P > 20^{d}$ with quality index $QI$ of 3 or better.  
Open circles are for the remainder of the variables in
Table~\ref{tbl4}. The drawn lines are the adopted P-L relations from
Madore \& Freedman (1991) in
equations (1) and (2) used also in the previous papers of this series.  
The ridge lines have been put arbitrarily at a 
modulus of
$(m -M)_{V} = 31.05$ and $(m-M)_{I} = 30.83$ as a first estimate, 
before analysis of differential extinction. The dashed upper and lower 
parallel lines indicate the expected projection of the width of the 
instability strip. The heavy spillage of points outside these bounds indicates
the presence of large differential extinction or noise or both. 
\label{fig6}}

\figcaption[]{Diagnostic diagram for the detection of differential
reddening, described in the text. The top panel shows the $V$-modulus 
on the abscissa. The lower panel is the same figure, but with 
the $I$-modulus on the abscissa.  In addition, on the lower panel, 
objects from Chip~4 are marked by a concentric outer circle.
\label{fig7}}

\figcaption[]{Diagram showing the $I$-modulus of individual Cepheids plotted 
against the $V$-modulus. Filled circles are for Cepheids with periods between 
20 and 60 days, and QI of 3 or higher. The solid line shows the reddening 
vector, and the dashed line is the slope suggested by the data, indicating 
that the spread in modulus values is not dominated by differential reddening,
but is due to additional correlated errors in photometry.
\label{fig8}}

\figcaption[]{The color-magnitude diagrams shown separately for the 4 
different chips. Note the abnormal appearance of the panel from 
chip 4, indicating problems with the photometry (see text).
\label{fig9}}

\figcaption[]{Same as Fig.~9, but with $F814W$ mags as the ordinate.
\label{fig10}}

\figcaption[]{The color-magnitude diagram shown for the 
regions in the different quadrants for the field in Chip 4 only.
Note that the abnormal appearance is in the two left hand quadrants
where the surface brightness, dust-lane appearance, and crowding are 
most pronounced. The CMD from the 
sparser right hand section of the Chip 4 field 
shows no abnormality.
\label{fig11}}

\figcaption[]{The Period-Color relation for the Cepheids in NGC~4527 is shown.
The filled circles show Cepheids with $ 60^{d} > P > 20^{d}$ with quality 
index $QI$ of 3 or better. Objects on Chip~4 are shown with a concentric 
outer circle. Note again how many of the Chip~4 objects show up as 
outliers. The continuous and dashed 
lines show the fiducial mean P-C relations (from 2 
different sources -- see text). Since most of the points are redder than the 
fiducial relation, there is clearly reddening present. The dotted lines 
show the color cut made (see text)  which run parallel to the fiducial 
P-C relation. 
\label{fig12}}

\figcaption[]{The P-L relation shows all Cepheids as in Fig.~6, but with the 
filled circles now showing only those objects with derived periods between 
20 and 65 days, quality index 3 or higher, and only those objects on 
Chip 4 which lie on the half of the chip where $X > 400$. The lines 
show canonical P-L relations corresponding to 
the final adopted DoPHOT distance moduli.
\label{fig12plus}}

\figcaption[]{Color-magnitude diagram $V-I$ versus $V$ obtained with ROMAFOT
photometry. Cepheids are indicated as filled circles.
\label{fig13}}

\figcaption{Color-magnitude diagram $V-I$ versus $I$ obtained with
ROMAFOT photometry. Cepheids are indicated as filled circles.
\label{fig14}}

\figcaption{Comparison of the ROMAFOT and DoPHOT photometry for all chip
and filter combinations. The horizontal line marks the average
difference using stars brighter than $25^{\rm m}$ 
in $V$ and $24^{\rm m}$ in $I$.
\label{fig15}}

\figcaption{Comparison of the ROMAFOT and DoPHOT photometry for all Cepheid
candidates given in Table~\ref{tbl5} for $V$ and $I$. Open
symbols represents $I$ and filled circles $V$ measurements.
\label{fig16}}

\figcaption{Diagnostic diagram to detect the presence of differential
absorption. Filled circles indicate Cepheids with $20 \leq P \leq 65$ and 
quality index 3 or better; open circles show the remaining candidates.
The candidates C1-V11, C2-V6, C3-V12 and C4-V6 are labeled.
\label{fig17}}

\figcaption{Period-luminosity relation in $V$ (top) and $I$ (bottom) based on
the ROMAFOT photometry. Filled circles indicate Cepheids with 
$20 \leq P \leq 65$ and quality index
3 or better; open circles show the remaining candidates.
The candidates C1-V11, C2-V6, C3-V12 and C4-V6 are labeled.
\label{fig18}}

\clearpage
\setcounter{table}{0}
\begin{deluxetable}{lccc}
\tablecaption{Journal of Observations. \label{tbl1}}
\tablewidth{0pt} 
\tablehead{
\colhead{Data Archive Designation} &\colhead{HJD at Midexposure} &
\colhead{Filter} & \colhead{Exposure time (s)}}
\startdata
u42g0101r + ...02r & 2451279.84066 & $F555W$ & 2500 \nl
u42g0103r + ...04r & 2451279.97330 & $F814W$ & 2500 \nl 
u42g0201r + ...02r & 2451288.17122 & $F555W$ & 2500 \nl
u42g0301r + ...02r & 2451295.29066 & $F555W$ & 2500 \nl
u42g0303r + ...04r & 2451295.42296 & $F814W$ & 2500 \nl
u42g0401r + ...02r & 2451301.20074 & $F555W$ & 2500 \nl
u42g0501r + ...02r & 2451341.83407 & $F555W$ & 2500 \nl
u42g0503r + ...04r & 2451341.96740 & $F814W$ & 2500 \nl
u42g0601r + ...02r & 2451311.00733 & $F555W$ & 2500 \nl
u42g0603r + ...04r & 2451311.13928 & $F814W$ & 2500 \nl
u42g0701r + ...02r & 2451315.03824 & $F555W$ & 2500 \nl
u42g0801r + ...02r & 2451319.67260 & $F555W$ & 2500 \nl
u42g0901r + ...02r & 2451325.18059 & $F555W$ & 2500 \nl
u42g1001r + ...02r & 2451331.96185 & $F555W$ & 2500 \nl
u42g1003r + ...04r & 2451332.09553 & $F814W$ & 2500 \nl
u42g1101r + ...02r & 2451339.81879 & $F555W$ & 2500 \nl
u42g1201r + ...02m & 2451348.81809 & $F555W$ & 2500 \nl
\enddata
\end{deluxetable}
\newpage
\begin{deluxetable}{lcclcc}
\footnotesize
\tablecaption{Position of the Variable Stars on {\it U42G0101R~\&~02R}
\label{tbl2}}
\tablewidth{0pt}
\tablehead{
\colhead{Variable ID} & \colhead{X-position} & \colhead{Y-position} &
\colhead{\quad\quad Variable ID} & \colhead{X-position} & 
\colhead{Y-position}} 
\startdata
C1-V1 & 148.36 & 460.08 & \quad\quad C1-V2 & 298.30 & 402.49 \nl
C1-V3 & 360.12 & 407.32 & \quad\quad C1-V4 & 375.35 & 236.14 \nl
C1-V5 & 403.07 & 521.44 & \quad\quad C1-V6 & 410.44 & ~97.22 \nl
C1-V7 & 484.49 & 489.44 & \quad\quad C1-V8 & 492.13 & 678.84 \nl
C1-V9 & 522.73 & 548.80 & \quad\quad C1-V10 & 527.21 & 301.16 \nl
C1-V11 & 540.98 & 347.19 & \quad\quad C1-V12 & 554.97 & 482.63 \nl
C1-V13 & 589.86 & 126.94 & \quad\quad C1-V14 & 603.22 & 462.17 \nl
C1-V15 & 618.72 & ~85.29 & \quad\quad C1-V16 & 623.71 & 102.30 \nl
C1-V17 & 742.17 & 131.15 & \quad\quad C1-V18 & 795.62 & 439.90 \nl
C2-V1 & ~82.70 & 259.27 & \quad\quad C2-V2 & 127.69 & 275.40 \nl
C2-V3 & 237.64 & 559.45 & \quad\quad C2-V4 & 270.90 & 102.82 \nl
C2-V5 & 275.83 & 258.40 & \quad\quad C2-V6 & 542.05 & 240.74 \nl 
C3-V1 & 129.36 & 528.00 & \quad\quad C3-V2 & 134.03 & 387.03 \nl
C3-V3 & 135.17 & 282.31 & \quad\quad C3-V4 & 194.25 & 274.95 \nl
C3-V5 & 300.00 & 341.11 & \quad\quad C3-V6 & 302.60 & 267.74 \nl
C3-V7 & 305.61 & 255.01 & \quad\quad C3-V8 & 359.56 & 255.11 \nl
C3-V9 & 369.25 & 264.58 & \quad\quad C3-V10 & 436.53 & 259.57 \nl
C3-V11 & 454.94 & 274.09 & \quad\quad C3-V12 & 586.75 & 178.44 \nl
C3-V13 & 672.26 & 168.40 & \quad\quad C3-V14 & 701.09 & 196.39 \nl
C3-V15 & 733.89 & 444.13 & \quad\quad C3-V16 & 782.02 & 326.09 \nl
C4-V1 & ~76.79 & 575.46 & \quad\quad C4-V2 & ~90.05 & 352.98 \nl
C4-V3 & 125.86 & 248.07 & \quad\quad C4-V4 & 136.55 & 176.03 \nl
C4-V5 & 186.83 & 108.56 & \quad\quad C4-V6 & 209.69 & 472.61 \nl
C4-V7 & 249.80 & 126.75 & \quad\quad C4-V8 & 253.50 & 283.67 \nl
C4-V9 & 262.62 & 553.92 & \quad\quad C4-V10 & 279.05 & ~98.59 \nl
C4-V11 & 328.98 & 314.79 & \quad\quad C4-V12 & 351.20 & 717.17 \nl
C4-V13 & 359.12 & 608.59 & \quad\quad C4-V14 & 359.96 & 200.73 \nl
C4-V15 & 384.24 & 435.91 & \quad\quad C4-V16 & 391.35 & 669.77 \nl
C4-V17 & 397.16 & 337.02 & \quad\quad C4-V18 & 405.67 & 417.76 \nl
C4-V19 & 410.94 & 344.78 & \quad\quad C4-V20 & 445.35 & 187.90 \nl
C4-V21 & 456.87 & 757.15 & \quad\quad C4-V22 & 462.33 & 769.18 \nl
C4-V23 & 508.54 & 169.13 & \quad\quad C4-V24 & 510.38 & 569.40 \nl
C4-V25 & 535.82 & 201.03 & \quad\quad C4-V26 & 558.990 & 539.71 \nl
\enddata
\end{deluxetable}
\newpage
\begin{deluxetable}{lcccccccc}
\tablecaption{Photometry of Variable Stars: Magnitudes and Error Estimates
\label{tbl3}}
\tiny
\tablewidth{0pt}
\tablehead{ }
\startdata
\tableline\tableline
 HJD & C1-V1 & C1-V2 & C1-V3 & C1-V4 & C1-V5 & C1-V6 & C1-V7 & C1-V8 \nl
\tableline
\multicolumn{9}{c}{$F555W$} \nl
\tableline
 2451331.9619 & 25.27  0.06 & 26.42  0.14 & 25.26  0.06 & 26.66  0.15 & 25.43  
0.06 & 26.43  0.17 & 24.84  0.05 & 25.94  0.07  \nl
 2451339.8188 & 25.38  0.08 & 26.55  0.16 & 25.34  0.06 & 25.89  0.10 & 25.93  
0.09 & 25.68  0.09 & 24.94  0.05 & 26.58  0.44  \nl
 2451348.8181 & 25.26  0.06 & 26.06  0.10 & 25.31  0.05 & 26.61  0.15 & 26.08  
0.16 & 25.66  0.08 & 25.22  0.05 & 25.72  0.07  \nl
 2451279.8407 & 24.93  0.05 & 26.56  0.14 & 25.52  0.06 & 26.05  0.12 & 25.44  
0.06 & 26.27  0.13 & 24.88  0.05 & 26.52  0.17  \nl
 2451288.1712 & 24.86  0.05 & 26.14  0.09 & 25.38  0.06 & 26.31  0.11 & 25.81  
0.08 & 25.73  0.08 & 24.96  0.05 & 25.93  0.08  \nl
 2451295.2907 & 24.80  0.09 & 26.84  0.24 & 25.45  0.06 & 25.48  0.09 & 26.24  
0.12 & 25.88  0.09 & 25.14  0.05 & 26.43  0.10  \nl
 2451301.2007 & 24.85  0.05 & 26.18  0.09 & 24.95  0.08 & 26.08  0.15 & 25.11  
0.07 & 25.55  0.07 & 25.30  0.06 & 26.23  0.12  \nl
 2451341.8341 & 25.32  0.09 & 26.47  0.14 &     ---     & 25.95  0.10 & 26.12  
0.11 & 25.93  0.11 & 25.02  0.07 & 26.27  0.11  \nl
 2451311.0073 & 24.94  0.05 & 26.39  0.15 & 25.21  0.06 & 26.51  0.13 & 25.64  
0.09 & 25.51  0.09 & 25.49  0.07 & 25.98  0.07  \nl
 2451315.0382 & 25.05  0.06 & 26.71  0.15 & 25.01  0.08 & 25.45  0.08 & 26.13  
0.10 & 25.45  0.08 & 25.46  0.06 & 26.21  0.08  \nl
 2451319.6726 & 25.06  0.05 & 26.66  0.16 & 25.47  0.06 & 25.94  0.10 & 26.36  
0.13 & 25.97  0.11 & 25.15  0.05 & 26.38  0.13  \nl
 2451325.1806 & 25.16  0.05 & 25.94  0.08 & 25.31  0.05 & 26.68  0.16 & 26.13  
0.11 & 25.60  0.08 & 24.68  0.04 & 25.65  0.08  \nl
\tableline
\multicolumn{9}{c}{$F814W$} \nl
\tableline
 2451332.0955 & 23.78  0.05 & 25.31  0.12 & 24.86  0.11 & 25.55  0.20 & 23.96  
0.06 & 25.99  0.23 & 23.76  0.05 & 24.90  0.08  \nl
 2451279.9733 & 23.63  0.05 & 25.53  0.16 & 25.14  0.12 & 24.98  0.12 & 24.49  
0.05 & 25.61  0.17 & 24.03  0.05 & 25.20  0.11  \nl
 2451295.4230 & 23.47  0.05 & 25.63  0.14 & 25.01  0.13 & 24.92  0.12 & 25.05  
0.13 & 25.86  0.43 & 23.88  0.06 & 24.92  0.12  \nl
 2451341.9674 & 23.82  0.05 & 25.33  0.14 & 25.00  0.11 & 24.94  0.12 & 24.85  
0.09 & 25.33  0.15 & 23.64  0.05 & 25.47  0.13  \nl
 2451311.1393 & 23.55  0.05 & 25.17  0.12 & 25.17  0.11 & 25.71  0.19 & 24.62  
0.08 & 25.23  0.14 & 24.38  0.06 & 25.11  0.13  \nl
\tableline\tableline
  HJD & C1-V9 & C1-V10 & C1-V11 & C1-V12 & C1-V13 & C1-V14 & C1-V15 & C1-V16 \nl
\tableline
\multicolumn{9}{c}{$F555W$} \nl
\tableline    
 2451331.9619 & 27.12  0.20 & 25.06  0.06 & 27.35  0.24 & 25.97  0.17 & 25.22  
0.09 & 26.80  0.19 & 25.75  0.09 & 26.76  0.20  \nl
 2451339.8188 & 26.25  0.12 & 25.34  0.08 & 26.44  0.15 & 26.28  0.21 & 25.55  
0.14 & 26.18  0.09 & 26.83  0.26 & 25.85  0.18  \nl
 2451348.8181 & 26.72  0.19 & 25.63  0.09 & 26.73  0.28 & 25.38  0.22 &     ---  
   & 27.18  0.21 & 25.74  0.09 & 26.89  0.40  \nl
 2451279.8407 & 26.13  0.12 & 24.98  0.06 & 26.48  0.13 & 24.97  0.22 & 25.82  
0.16 & 27.20  0.28 & 26.63  0.20 & 26.82  0.23  \nl
 2451288.1712 & 27.01  0.28 & 25.12  0.06 & 27.17  0.21 & 25.45  0.07 & 25.86  
0.16 & 26.27  0.10 & 26.11  0.16 & 26.55  0.18  \nl
 2451295.2907 & 27.01  0.20 & 25.27  0.11 & 27.58  0.39 & 25.83  0.10 & 25.73  
0.37 & 26.97  0.19 & 26.38  0.19 & 26.12  0.12  \nl
 2451301.2007 & 26.50  0.13 & 25.46  0.09 & 26.70  0.16 & 25.90  0.11 & 24.92  
0.12 & 26.32  0.10 & 25.97  0.13 & 26.06  0.10  \nl
 2451341.8341 & 26.32  0.16 & 25.41  0.09 & 26.33  0.14 & 25.94  0.11 & 25.05  
0.09 & 26.52  0.15 & 26.35  0.19 & 26.23  0.18  \nl
 2451311.0073 & 27.00  0.21 & 25.09  0.06 & 26.56  0.16 & 25.83  0.13 & 25.48  
0.15 & 26.69  0.18 & 26.36  0.16 & 26.88  0.28  \nl
 2451315.0382 & 26.38  0.14 & 24.82  0.05 & 26.76  0.15 & 25.56  0.11 & 25.02  
0.09 & 26.44  0.12 & 25.64  0.09 & 25.83  0.10  \nl
 2451319.6726 & 26.72  0.17 & 24.80  0.05 & 27.10  0.35 & 25.38  0.19 & 26.38  
0.45 & 27.31  0.27 & 25.99  0.12 & 25.97  0.15  \nl
 2451325.1806 & 26.80  0.19 & 24.96  0.05 & 27.82  0.39 & 25.22  0.09 & 25.62  
0.12 & 26.95  0.21 & 26.91  0.25 & 26.15  0.13  \nl
\tableline
\multicolumn{9}{c}{$F814W$} \nl
\tableline
 2451332.0955 & 25.80  0.22 & 23.99  0.06 & 25.24  0.14 &     ---     & 24.69  
0.15 & 25.81  0.17 & 25.22  0.16 & 25.55  0.21  \nl
 2451279.9733 &     ---     & 23.95  0.05 & 24.97  0.13 & 24.35  0.09 &     ---  
   & 26.07  0.22 & 25.75  0.22 & 25.22  0.15  \nl
 2451295.4230 &     ---     & 24.17  0.09 & 25.52  0.17 &     ---     & 25.09  
0.45 & 26.12  0.25 & 25.17  0.14 & 25.15  0.17  \nl
 2451341.9674 & 25.64  0.16 & 24.20  0.07 & 24.72  0.09 &     ---     & 25.88  
0.28 & 25.71  0.18 & 25.64  0.20 & 25.53  0.22  \nl
 2451311.1393 &     ---     & 23.98  0.08 & 24.86  0.09 &     ---     & 24.13  
0.11 & 25.40  0.15 & 25.64  0.25 & 26.38  0.42  \nl
\tableline\tableline
  HJD & C1-V17 & C1-V18 & C2-V1 & C2-V2 & C2-V3 & C2-V4 & C2-V5 & C2-V6 \nl
\tableline
\multicolumn{9}{c}{$F555W$} \nl
\tableline    
 2451331.9619 & 27.14  0.31 & 26.57  0.17 & 25.74  0.09 & 26.07  0.12 & 25.80  
0.06 & 27.51  0.35 & 27.07  0.21 & 26.49  0.12  \nl
 2451339.8188 & 25.43  0.09 & 27.01  0.28 & 25.49  0.07 & 26.45  0.36 & 26.65  
0.13 & 27.34  0.30 & 27.65  0.42 & 26.75  0.16  \nl
 2451348.8181 & 26.07  0.14 & 26.17  0.13 & 25.05  0.06 & 25.36  0.09 & 25.73  
0.07 & 26.63  0.17 &     ---     & 26.75  0.16  \nl
 2451279.8407 & 25.42  0.11 & 27.32  0.43 & 25.39  0.07 & 25.99  0.12 & 26.23  
0.09 & 26.62  0.15 &     ---     & 27.27  0.25  \nl
 2451288.1712 & 25.77  0.12 & 26.52  0.12 & 25.71  0.08 & 26.22  0.13 & 26.34  
0.11 & 27.06  0.21 & 26.01  0.09 & 27.39  0.26  \nl
 2451295.2907 & 25.99  0.24 & 26.70  0.22 & 25.70  0.09 & 25.48  0.09 & 26.66  
0.12 & 26.96  0.19 & 26.48  0.14 & 27.50  0.31  \nl
 2451301.2007 & 26.81  0.31 & 27.42  0.45 & 25.30  0.07 & 25.80  0.10 & 26.07  
0.27 & 27.55  0.33 & 26.62  0.14 & 27.64  0.38  \nl
 2451341.8341 & 25.70  0.12 & 26.34  0.16 & 24.97  0.06 & 26.61  0.24 & 26.42  
0.11 & 27.81  0.49 & 27.52  0.36 & 26.71  0.14  \nl
 2451311.0073 & 25.36  0.08 & 25.77  0.18 & 25.24  0.06 &     ---     & 25.73  
0.06 & 27.45  0.27 & 26.58  0.13 & 26.49  0.09  \nl
 2451315.0382 & 25.99  0.16 & 26.84  0.22 & 25.40  0.06 & 25.69  0.09 & 26.25  
0.09 & 25.56  0.09 & 27.28  0.46 & 26.38  0.12  \nl
 2451319.6726 & 26.12  0.14 & 26.40  0.12 & 25.55  0.08 & 25.56  0.10 & 26.71  
0.13 & 27.73  0.44 & 27.14  0.26 & 26.42  0.11  \nl
 2451325.1806 & 26.34  0.16 & 26.45  0.17 &  0.00  0.00 &  0.00  0.00 &  0.00  
0.00 &  0.00  0.00 &  0.00  0.00 &  0.00  0.00  \nl
\tableline
\multicolumn{9}{c}{$F814W$} \nl
\tableline
 2451332.0955 & 25.63  0.24 & 26.28  0.33 & 24.68  0.09 & 24.98  0.12 & 25.08  
0.10 &     ---     & 26.12  0.22 & 25.13  0.10  \nl
 2451279.9733 & 24.52  0.10 & 25.53  0.22 & 23.96  0.08 & 24.91  0.14 & 25.30  
0.12 &     ---     & 25.30  0.15 & 25.77  0.14  \nl
 2451295.4230 & 24.76  0.14 & 25.63  0.19 & 24.81  0.12 & 24.67  0.13 & 25.41  
0.14 &     ---     & 25.66  0.20 & 26.28  0.32  \nl
 2451341.9674 & 24.69  0.11 & 26.07  0.23 & 24.13  0.07 & 25.34  0.15 & 25.56  
0.13 &     ---     & 26.15  0.25 & 25.49  0.12  \nl
 2451311.1393 & 24.56  0.09 & 26.22  0.40 & 24.09  0.07 & 25.26  0.14 & 25.39  
0.11 &     ---     & 25.69  0.16 & 25.71  0.16  \nl
\tableline\tableline     
 HJD & C3-V1 & C3-V2 & C3-V3 & C3-V4 & C3-V5 & C3-V6 & C3-V7 & C3-V8 \nl
\tableline
\multicolumn{9}{c}{$F555W$} \nl
\tableline      
 2451331.9619 & 26.15  0.13 & 25.62  0.09 & 27.56  0.43 &     ---     & 25.11  
0.05 & 26.26  0.17 & 24.80  0.04 & 24.98  0.04  \nl
 2451339.8188 & 26.72  0.22 & 24.88  0.05 & 25.92  0.10 & 25.26  0.05 & 25.33  
0.07 & 25.96  0.11 & 24.55  0.04 & 25.15  0.06  \nl
 2451348.8181 & 25.81  0.11 & 25.14  0.07 & 27.05  0.27 & 25.39  0.08 & 25.31  
0.07 & 25.66  0.11 & 24.72  0.04 & 25.25  0.07  \nl
 2451279.8407 & 26.63  0.21 & 25.46  0.09 & 26.50  0.18 & 25.35  0.06 & 25.10  
0.05 & 26.21  0.18 & 24.65  0.07 & 24.98  0.06  \nl
 2451288.1712 & 25.58  0.09 & 26.02  0.17 & 26.91  0.22 & 25.47  0.07 & 25.21  
0.06 & 25.63  0.10 & 24.80  0.06 & 25.21  0.06  \nl
 2451295.2907 & 26.20  0.13 & 25.48  0.09 & 26.40  0.22 & 25.18  0.07 & 25.35  
0.07 & 25.68  0.10 & 24.88  0.05 & 25.28  0.05  \nl
 2451301.2007 & 26.60  0.17 & 24.84  0.05 & 26.76  0.22 & 24.93  0.05 & 25.43  
0.06 & 25.69  0.10 & 24.92  0.05 & 25.42  0.07  \nl
 2451341.8341 & 26.71  0.22 & 24.88  0.05 & 25.49  0.08 & 25.27  0.07 &     ---  
   & 25.30  0.07 & 24.67  0.08 & 25.14  0.06  \nl
 2451311.0073 & 25.99  0.11 & 25.25  0.07 & 26.42  0.15 & 24.93  0.06 & 25.20  
0.08 & 26.35  0.16 & 24.95  0.05 & 25.13  0.06  \nl
 2451315.0382 & 26.22  0.14 & 25.36  0.07 & 26.49  0.16 & 24.86  0.05 & 25.11  
0.06 & 25.35  0.06 & 24.98  0.05 & 24.92  0.06  \nl
 2451319.6726 & 26.95  0.22 & 25.68  0.10 & 26.80  0.25 & 24.93  0.05 & 25.45  
0.07 & 25.53  0.09 & 24.90  0.05 & 24.95  0.06  \nl
 2451325.1806 & 26.02  0.11 & 25.81  0.10 & 26.22  0.18 & 25.07  0.06 & 25.38  
0.07 & 25.93  0.09 & 24.92  0.05 & 24.94  0.05  \nl
\tableline
\multicolumn{9}{c}{$F814W$} \nl
\tableline
 2451332.0950 & 25.05  0.12 & 25.13  0.30 & 26.11  0.29 & 23.76  0.05 & 23.53  
0.07 & 24.43  0.09 & 24.26  0.06 & 23.77  0.05  \nl
 2451279.9730 & 25.69  0.25 & 24.38  0.08 & 25.52  0.17 & 23.98  0.05 & 24.34  
0.09 & 24.69  0.08 & 24.09  0.05 & 23.78  0.05  \nl
 2451295.4220 & 25.10  0.14 & 24.61  0.09 & 25.83  0.26 & 23.85  0.06 & 24.58  
0.10 & 24.34  0.08 & 24.41  0.08 & 24.07  0.05  \nl
 2451341.9670 & 25.80  0.22 & 24.13  0.06 & 26.38  0.39 & 23.90  0.07 & 24.36  
0.10 & 24.23  0.07 & 24.06  0.06 & 23.92  0.05  \nl
 2451311.1390 & 25.05  0.11 & 24.09  0.06 & 25.75  0.20 & 23.28  0.05 & 24.40  
0.08 & 24.64  0.09 & 24.49  0.08 & 23.91  0.05  \nl
\tableline\tableline     
 HJD & C3-V9 & C3-V10 & C3-V11 & C3-V12 & C3-V13 & C3-V14 & C3-V15 & C3-V16 \nl
\tableline
\multicolumn{9}{c}{$F555W$} \nl
\tableline       
 2451331.9619 & 24.48  0.04 & 26.06  0.12 & 25.78  0.11 & 26.06  0.14 & 25.08  
0.08 & 26.65  0.16 & 26.51  0.15 & 25.83  0.08  \nl
 2451339.8188 & 24.60  0.05 & 25.66  0.09 & 26.63  0.23 & 25.75  0.08 & 25.00  
0.06 & 25.62  0.18 & 26.80  0.18 & 26.36  0.16  \nl
 2451348.8181 & 24.80  0.06 & 25.95  0.11 & 25.97  0.13 & 25.38  0.08 & 24.71  
0.06 & 25.89  0.16 & 26.11  0.10 & 25.28  0.17  \nl
 2451279.8407 & 24.32  0.05 & 25.99  0.12 & 26.83  0.27 & 25.67  0.11 & 24.77  
0.06 & 26.10  0.14 & 25.84  0.10 & 25.21  0.08  \nl
 2451288.1712 & 24.58  0.08 & 26.13  0.12 & 26.03  0.13 & 25.59  0.08 & 24.72  
0.05 & 26.30  0.12 & 27.02  0.21 & 26.15  0.11  \nl
 2451295.2907 & 24.86  0.05 & 25.94  0.10 & 27.05  0.30 & 25.30  0.08 & 24.74  
0.06 & 25.57  0.08 & 26.40  0.15 & 26.43  0.16  \nl
 2451301.2007 & 24.95  0.06 & 25.55  0.08 & 25.83  0.10 & 25.55  0.10 & 24.85  
0.06 & 26.05  0.14 & 26.98  0.18 & 25.18  0.05  \nl
 2451341.8341 & 24.76  0.05 & 25.75  0.11 & 27.44  0.38 & 25.43  0.08 & 24.84  
0.04 & 25.34  0.06 & 26.97  0.23 & 26.39  0.19  \nl
 2451311.0073 & 24.99  0.05 & 25.56  0.09 & 26.83  0.21 & 25.65  0.10 & 25.02  
0.05 & 26.60  0.29 & 26.66  0.18 & 26.00  0.10  \nl
 2451315.0382 & 24.71  0.05 & 25.80  0.08 & 26.77  0.19 & 25.74  0.10 & 25.08  
0.06 & 25.80  0.08 & 27.00  0.22 & 26.12  0.09  \nl
 2451319.6726 & 24.41  0.05 & 25.43  0.09 & 26.24  0.14 & 26.01  0.13 & 25.08  
0.06 & 25.53  0.15 & 25.85  0.10 & 26.38  0.13  \nl
 2451325.1806 & 24.42  0.04 & 25.65  0.09 & 26.89  0.23 & 25.91  0.12 & 25.12  
0.06 & 26.02  0.11 & 26.64  0.16 & 25.41  0.06  \nl
\tableline
\multicolumn{9}{c}{$F814W$} \nl
\tableline
 2451332.0950 & 23.40  0.06 & 24.69  0.10 & 24.80  0.10 & 24.23  0.09 & 23.76  
0.05 & 25.31  0.14 & 25.74  0.17 & 24.75  0.08  \nl
 2451279.9730 & 23.52  0.05 & 24.99  0.11 & 26.11  0.31 & 24.39  0.09 & 23.67  
0.05 & 25.11  0.11 & 25.44  0.13 & 24.63  0.07  \nl
 2451295.4220 & 23.88  0.05 & 24.81  0.09 & 26.11  0.31 & 24.16  0.11 & 23.56  
0.05 & 24.85  0.10 & 25.66  0.18 & 25.30  0.14  \nl
 2451341.9670 & 23.30  0.06 & 25.32  0.25 & 25.88  0.21 & 24.17  0.08 & 23.66  
0.05 & 24.78  0.16 & 26.09  0.22 & 25.57  0.23  \nl
 2451311.1390 & 23.82  0.05 & 24.57  0.09 & 25.84  0.22 & 24.03  0.09 & 23.65  
0.05 & 25.35  0.17 & 25.12  0.11 & 25.10  0.09  \nl
\tableline\tableline     
 HJD &  C4-V1 & C4-V2 & C4-V3 & C4-V4 & C4-V5 & C4-V6 & C4-V7 & C4-V8 \nl  
\tableline
\multicolumn{9}{c}{$F555W$} \nl
\tableline      
 2451331.9619 & 25.44  0.10 & 25.30  0.10 & 26.05  0.22 & 24.77  0.10 & 26.42  
0.19 & 24.19  0.06 & 26.52  0.16 & 24.74  0.13  \nl
 2451339.8188 & 25.61  0.13 & 24.83  0.06 & 25.00  0.11 & 24.41  0.08 & 26.21  
0.15 & 24.30  0.06 & 27.70  0.47 & 24.67  0.12  \nl
 2451348.8181 & 25.05  0.08 & 25.03  0.08 & 25.35  0.12 & 24.62  0.09 & 25.60  
0.07 & 25.24  0.18 & 26.36  0.18 & 24.95  0.13  \nl
 2451279.8407 & 25.52  0.12 & 25.23  0.09 & 25.85  0.23 & 24.65  0.10 & 25.73  
0.16 & 24.60  0.07 & 26.93  0.26 & 24.77  0.14  \nl
 2451288.1712 & 24.76  0.06 & 25.41  0.15 & 25.45  0.15 & 24.57  0.10 & 26.16  
0.16 & 24.53  0.06 & 26.78  0.30 & 25.33  0.19  \nl
 2451295.2907 & 25.19  0.08 & 25.13  0.08 & 25.80  0.26 & 24.70  0.11 & 25.86  
0.11 & 24.00  0.04 & 25.60  0.09 & 25.18  0.16  \nl
 2451301.2007 & 25.52  0.10 & 24.86  0.06 & 25.85  0.21 & 24.82  0.12 & 26.25  
0.13 & 24.31  0.06 & 27.36  0.42 & 24.67  0.12  \nl
 2451341.8341 & 25.52  0.12 & 24.91  0.08 & 25.08  0.11 & 24.42  0.10 & 26.50  
0.19 & 24.31  0.06 & 27.38  0.43 & 25.19  0.12  \nl
 2451311.0073 & 25.56  0.12 & 24.99  0.08 & 25.18  0.14 & 25.03  0.12 & 26.12  
0.16 & 24.11  0.05 & 26.36  0.16 & 25.47  0.22  \nl
 2451315.0382 & 25.18  0.10 & 25.09  0.08 & 25.05  0.12 & 25.39  0.27 & 26.24  
0.16 & 24.13  0.07 & 25.64  0.13 & 24.80  0.15  \nl
 2451319.6726 & 25.05  0.08 & 25.23  0.09 & 25.61  0.17 & 25.18  0.15 & 25.77  
0.09 & 24.26  0.05 & 27.01  0.25 & 24.55  0.13  \nl
 2451325.1806 & 25.25  0.09 & 25.24  0.09 & 26.05  0.35 & 25.10  0.15 & 26.01  
0.12 & 24.59  0.08 & 27.31  0.34 & 25.24  0.13  \nl
\tableline
\multicolumn{9}{c}{$F814W$} \nl
\tableline
 2451332.0950 & 24.14  0.11 & 24.02  0.11 & 24.74  0.23 & 23.43  0.09 & 24.94  
0.13 & 23.13  0.15 &     ---     & 23.68  0.32  \nl
 2451279.9730 & 24.17  0.10 & 23.89  0.07 & 24.84  0.24 & 23.36  0.07 & 24.70  
0.12 & 23.30  0.22 &     ---     &     ---      \nl
 2451295.4220 & 23.75  0.11 & 24.08  0.10 & 24.49  0.14 & 23.31  0.06 & 24.94  
0.14 & 23.02  0.14 &     ---     & 23.65  0.31  \nl
 2451341.9670 & 24.59  0.22 & 23.57  0.07 & 24.41  0.16 & 22.91  0.09 & 25.07  
0.17 & 23.45  0.23 &     ---     & 24.00  0.21  \nl
 2451311.1390 & 24.64  0.16 & 23.45  0.10 & 24.25  0.15 & 23.68  0.10 & 25.02  
0.17 & 23.18  0.16 &     ---     &     ---      \nl
\tableline\tableline     
 HJD &  C4-V9 & C4-V10 & C4-V11 & C4-V12 & C4-V13 & C4-V14 & C4-V15 & C4-V16 \nl  
\tableline
\multicolumn{9}{c}{$F555W$} \nl
\tableline     
 2451331.9619 & 25.38  0.16 & 26.42  0.15 & 26.19  0.20 & 25.47  0.10 & 26.23  
0.21 & 26.26  0.13 & 26.05  0.16 & 25.80  0.11  \nl
 2451339.8188 & 25.31  0.16 & 26.30  0.14 & 25.65  0.12 & 25.56  0.12 & 24.98  
0.08 & 25.43  0.07 & 25.71  0.12 & 25.91  0.14  \nl
 2451348.8181 & 24.89  0.12 & 25.99  0.11 & 26.07  0.19 & 24.88  0.07 & 25.28  
0.09 & 25.69  0.09 & 26.14  0.16 & 24.99  0.07  \nl
 2451279.8407 & 25.19  0.12 & 25.76  0.09 & 26.06  0.19 & 25.08  0.14 & 25.55  
0.11 & 25.42  0.08 & 26.15  0.17 & 25.00  0.06  \nl
 2451288.1712 & 24.92  0.11 & 26.99  0.29 & 25.75  0.14 & 25.35  0.10 & 25.83  
0.14 & 25.99  0.11 & 25.40  0.08 & 25.35  0.08  \nl
 2451295.2907 & 25.24  0.12 & 25.98  0.14 & 26.34  0.23 & 25.30  0.12 & 26.25  
0.21 & 26.23  0.14 & 25.45  0.11 & 25.32  0.13  \nl
 2451301.2007 & 25.41  0.14 &     ---     & 25.53  0.10 & 25.40  0.12 & 25.05  
0.08 & 26.22  0.16 & 25.97  0.14 & 25.94  0.14  \nl
 2451341.8341 & 25.39  0.16 & 26.82  0.32 & 25.68  0.13 & 25.56  0.15 & 25.02  
0.07 & 25.46  0.09 & 25.92  0.15 & 25.65  0.12  \nl
 2451311.0073 & 25.32  0.14 & 26.28  0.14 & 26.08  0.18 & 24.83  0.06 & 25.38  
0.11 & 25.60  0.08 & 25.50  0.09 & 25.26  0.13  \nl
 2451315.0382 & 24.97  0.10 & 26.96  0.21 & 26.39  0.25 & 24.85  0.08 & 25.44  
0.11 & 25.70  0.09 & 25.44  0.14 & 25.00  0.06  \nl
 2451319.6726 & 24.66  0.09 & 26.13  0.10 & 25.65  0.11 & 25.20  0.12 & 25.63  
0.13 & 25.94  0.13 & 25.84  0.12 & 25.26  0.07  \nl
 2451325.1806 & 25.07  0.12 & 26.58  0.17 & 25.74  0.12 & 25.33  0.08 & 25.83  
0.14 & 25.92  0.12 & 25.97  0.16 & 25.55  0.12  \nl
\tableline
\multicolumn{9}{c}{$F814W$} \nl
\tableline
 2451332.0950 & 23.84  0.13 & 25.26  0.19 &     ---     & 24.24  0.11 & 24.17  
0.12 & 24.51  0.12 & 24.99  0.18 & 24.26  0.11  \nl
 2451279.9730 & 24.11  0.19 & 24.71  0.14 &     ---     & 24.02  0.10 & 24.13  
0.15 & 24.02  0.07 & 25.12  0.22 & 23.55  0.08  \nl
 2451295.4220 & 23.81  0.19 & 24.90  0.17 &     ---     & 24.07  0.12 & 24.38  
0.17 & 24.52  0.13 & 24.88  0.17 & 24.74  0.16  \nl
 2451341.9670 & 23.89  0.21 & 25.18  0.16 &     ---     & 24.31  0.14 & 23.80  
0.08 & 24.07  0.08 & 24.80  0.16 & 24.33  0.11  \nl
 2451311.1390 & 23.94  0.16 & 24.78  0.14 &     ---     & 23.90  0.09 & 24.05  
0.12 & 24.01  0.08 & 24.55  0.16 & 24.03  0.10  \nl
\tableline\tableline     
 HJD &  C4-V17 & C4-V18 & C4-V19 & C4-V20 & C4-V21 & C4-V22 & C4-V23 & C4-V24 
\nl  
\tableline
\multicolumn{9}{c}{$F555W$} \nl
\tableline     
 2451331.9619 & 24.67  0.12 & 25.78  0.07 & 25.99  0.13 & 25.18  0.05 & 25.72  
0.13 & 24.60  0.06 & 25.09  0.05 & 25.30  0.07  \nl
 2451339.8188 & 26.01  0.19 & 26.12  0.17 & 26.67  0.26 & 25.30  0.06 & 25.86  
0.14 & 24.74  0.07 & 25.22  0.05 & 25.80  0.09  \nl
 2451348.8181 & 25.39  0.11 & 26.22  0.14 & 26.56  0.17 & 25.67  0.07 & 25.08  
0.07 & 24.88  0.05 & 25.31  0.05 & 25.14  0.05  \nl
 2451279.8407 & 25.49  0.11 & 25.89  0.09 & 26.06  0.12 & 25.09  0.06 & 25.62  
0.11 & 24.95  0.07 & 25.26  0.05 & 25.32  0.07  \nl
 2451288.1712 & 25.25  0.06 & 26.43  0.14 & 26.61  0.16 & 24.74  0.07 & 25.65  
0.10 & 24.50  0.05 & 25.32  0.07 & 25.70  0.11  \nl
 2451295.2907 & 25.80  0.16 & 25.66  0.08 & 26.56  0.21 & 25.45  0.07 & 25.15  
0.09 & 24.65  0.06 & 25.30  0.06 & 25.23  0.06  \nl
 2451301.2007 & 25.26  0.11 & 25.59  0.07 & 25.63  0.09 & 25.93  0.17 & 25.40  
0.10 & 24.84  0.07 & 24.88  0.04 & 25.68  0.10  \nl
 2451341.8341 & 25.86  0.14 & 26.23  0.12 & 27.08  0.32 & 25.31  0.06 & 26.03  
0.15 & 24.82  0.07 & 25.30  0.05 & 25.76  0.10  \nl
 2451311.0073 & 25.60  0.10 & 26.07  0.12 & 26.72  0.26 & 25.88  0.13 & 26.03  
0.19 & 24.95  0.08 & 24.85  0.05 & 25.30  0.08  \nl
 2451315.0382 & 25.26  0.11 & 26.09  0.10 & 27.16  0.36 & 25.12  0.05 & 26.13  
0.18 & 25.14  0.08 & 24.85  0.04 & 25.31  0.08  \nl
 2451319.6726 & 25.61  0.10 & 26.15  0.13 & 26.59  0.19 & 24.85  0.05 & 25.77  
0.14 & 25.07  0.08 & 24.89  0.05 & 25.68  0.12  \nl
 2451325.1806 & 26.30  0.39 & 25.10  0.07 & 26.14  0.15 & 24.94  0.05 & 25.38  
0.10 & 24.73  0.06 & 25.08  0.04 & 25.80  0.10  \nl
\tableline
\multicolumn{9}{c}{$F814W$} \nl
\tableline
 2451332.0950 & 24.74  0.19 &     ---     & 24.64  0.09 & 23.84  0.07 & 24.74  
0.17 & 23.62  0.09 & 23.72  0.05 & 24.38  0.08  \nl
 2451279.9730 & 24.53  0.17 & 24.49  0.10 & 24.64  0.10 & 23.88  0.06 & 24.64  
0.13 & 23.73  0.09 & 23.83  0.05 & 24.41  0.09  \nl
 2451295.4220 &     ---     & 24.51  0.10 & 24.74  0.12 & 23.93  0.06 & 24.21  
0.10 & 23.64  0.07 & 23.65  0.08 & 24.38  0.09  \nl
 2451341.9670 &     ---     & 24.74  0.12 & 24.63  0.14 & 23.90  0.08 & 24.77  
0.16 & 23.74  0.07 & 23.77  0.05 & 24.61  0.09  \nl
 2451311.1390 & 24.73  0.19 & 24.59  0.11 & 24.71  0.11 & 24.32  0.07 & 24.64  
0.11 & 23.83  0.08 & 23.35  0.05 & 24.40  0.09  \nl
\tableline\tableline     
 HJD &  C4-V25 & C4-V26 &&&&& \nl  
\tableline
\multicolumn{9}{c}{$F555W$} \nl
\tableline    
 2451331.9619 & 26.08  0.11 & 25.67  0.08  \nl
 2451339.8188 & 27.19  0.29 & 26.42  0.17  \nl
 2451348.8181 & 26.71  0.28 & 26.43  0.14  \nl
 2451279.8407 & 26.38  0.14 & 26.51  0.15  \nl
 2451288.1712 & 25.99  0.12 & 25.66  0.08  \nl
 2451295.2907 & 26.60  0.24 & 26.28  0.20  \nl
 2451301.2007 & 26.44  0.16 & 26.86  0.22  \nl
 2451341.8341 & 26.88  0.20 & 26.66  0.15  \nl
 2451311.0073 & 26.12  0.11 & 25.69  0.10  \nl
 2451315.0382 & 26.71  0.17 & 26.08  0.10  \nl
 2451319.6726 & 26.26  0.12 & 26.76  0.21  \nl
 2451325.1806 & 26.50  0.16 & 26.55  0.19  \nl
\tableline
\multicolumn{9}{c}{$F814W$} \nl
\tableline
 2451332.0950 & 25.21  0.17 & 24.83  0.12  \nl
 2451279.9730 & 25.10  0.14 & 25.41  0.21  \nl
 2451295.4220 & 25.33  0.19 & 24.98  0.17  \nl
 2451341.9670 & 25.21  0.16 & 25.30  0.19  \nl
 2451311.1390 & 25.19  0.14 & 24.92  0.14  \nl
\enddata
\end{deluxetable}
\newpage
\begin{deluxetable}{lcccccccrcc}
\tablecaption{Characteristics of the Cepheids \label{tbl4}}
\scriptsize
\tablewidth{0pt}
\tablehead{
 \colhead{Object} & \colhead{Period} & \colhead{$\langle{V}\rangle$} &
 \colhead{$\sigma_{\langle{V}\rangle}$} & \colhead{$\langle{I}\rangle$} & 
 \colhead{$\sigma_{\langle{I}\rangle}$} & \colhead{$U_{V}$} & 
 \colhead{$U_{I}$} & \colhead{$U_{T}$} & \colhead{$\sigma_{U_{T}}$} & 
 \colhead{Quality} \\
 \colhead{} & \colhead{(days)} & \colhead{} & 
 \colhead{} & \colhead{} & \colhead{} &
 \colhead{} & \colhead{} & \colhead{} &
 \colhead{} & \colhead{Index} \\
 \colhead{(1)} & \colhead{(2)} & \colhead{(3)} &
 \colhead{(4)} & \colhead{(5)} & \colhead{(6)} &
 \colhead{(7)} & \colhead{(8)} & \colhead{(9)} &
 \colhead{(10)} & \colhead{(11)}
 } 
\startdata
C1-V1  & 83.400 & 25.030 & 0.059 & 23.580 & 0.057 & 31.733 & 31.269 & 30.606 & 
0.27  & 3 \nl   
C1-V2  & 20.800 & 26.340 & 0.141 & 25.440 & 0.194 & 31.378 & 31.284 & 31.149 & 
0.56  & 3 \nl   
C1-V3  & 14.800 & 25.260 & 0.065 & 25.035 & 0.122 & 29.890 & 30.426 & 31.192 & 
0.38  & 3 \nl   
C1-V4  & 20.500 & 26.019 & 0.126 & 25.139 & 0.105 & 31.040 & 30.963 & 30.854 & 
0.38  & 5 \nl   
C1-V5  & 25.400 & 25.668 & 0.101 & 24.450 & 0.257 & 30.945 & 30.558 & 30.006 & 
0.68  & 5 \nl  
C1-V6  & 12.900 & 25.775 & 0.101 & 25.465 & 0.255 & 30.240 & 30.674 & 31.293 & 
0.67  & 3 \nl   
C1-V7  & 51.500 & 25.127 & 0.057 & 23.678 & 0.353 & 31.252 & 30.726 & 29.976 & 
0.89  & 3 \nl   
C1-V8  & 20.400 & 26.087 & 0.165 & 24.954 & 0.248 & 31.102 & 30.771 & 30.299 & 
0.68  & 4 \nl   
C1-V9  & 12.100 & 26.614 & 0.180 & 25.819 & 1.444 & 31.003 & 30.942 & 30.856 & 
3.52  & 1 \nl   
C1-V10 & 45.000 & 25.100 & 0.070 & 24.023 & 0.055 & 31.063 & 30.892 & 30.649 & 
0.28  & 3 \nl   
C1-V11 & 33.500 & 26.844 & 0.249 & 25.154 & 0.208 & 32.453 & 31.631 & 30.456 & 
0.66  & 5 \nl   
C1-V12 & 31.8:  & 25.69  & --- & --- & --- & --- & --- & --- & --- & 0 \nl  
C1-V13 & 14.200 & 25.491 & 0.243 & 25.384 & 0.777 & 30.071 & 30.720 & 31.647 & 
1.93  & 3 \nl   
C1-V14 & 13.200 & 26.654 & 0.181 & 25.757 & 0.173 & 31.147 & 30.996 & 30.781 & 
0.54  & 4 \nl   
C1-V15 & 15.400 & 26.066 & 0.170 & 25.299 & 0.315 & 30.743 & 30.743 & 30.743 & 
0.83  & 3 \nl   
C1-V16 & 23.000 & 26.257 & 0.189 & 25.260 & 0.258 & 31.415 & 31.237 & 30.982 & 
0.72  & 2 \nl  
C1-V17 & 28.400 & 25.940 & 0.203 & 24.843 & 0.231 & 31.351 & 31.101 & 30.742 & 
0.67  & 2 \nl   
C1-V18 &  7.400 & 26.536 & 0.254 & 25.767 & 0.628 & 30.335 & 30.237 & 30.098 & 
1.58  & 3 \nl   
C2-V1  & 38.900 & 25.340 & 0.072 & 24.208 & 0.267 & 31.128 & 30.884 & 30.534 & 
0.69  & 3 \nl   
C2-V2  & 27.800 & 25.871 & 0.169 & 24.904 & 0.149 & 31.256 & 31.133 & 30.956 & 
0.49  & 5 \nl   
C2-V3  & 16.700 & 26.226 & 0.137 & 25.166 & 0.195 & 31.001 & 30.718 & 30.313 & 
0.56  & 3 \nl   
C2-V4  & 13.9   & 26.83  &  ---  &   ---  &  ---  &   ---  &   ---  &   ---  & 
---   & 0 \nl
C2-V5  & 66.200 & 26.702 & 0.264 & 26.132 & 0.908 & 33.127 & 33.514 & 34.066 & 
2.25  & 4 \nl   
C2-V6  & 90.000 & 26.851 & 0.221 & 25.519 & 0.280 & 33.645 & 33.309 & 32.830 & 
0.78  & 5 \nl   
C3-V1  & 19.500 & 26.224 & 0.159 & 25.145 & 0.228 & 31.184 & 30.903 & 30.501 & 
0.64  & 5 \nl   
C3-V2  & 39.200 & 25.416 & 0.097 & 24.325 & 0.323 & 31.214 & 31.010 & 30.720 & 
0.83  & 5 \nl   
C3-V3  & 13.900 & 26.533 & 0.210 & 26.066 & 0.424 & 31.219 & 31.518 & 31.946 & 
1.09  & 1 \nl   
C3-V4  & 62.900 & 25.171 & 0.059 & 23.787 & 0.056 & 31.535 & 31.101 & 30.481 & 
0.27  & 2 \nl   
C3-V5  & 21.200 & 25.314 & 0.062 & 24.082 & 0.279 & 30.493 & 30.081 & 29.493 & 
0.72  & 1 \nl   
C3-V6  & 26.100 & 25.866 & 0.126 & 24.414 & 0.151 & 31.176 & 30.558 & 29.677 & 
0.46  & 2 \nl   
C3-V7  & 58.000 & 24.902 & 0.052 & 24.187 & 0.150 & 31.169 & 31.393 & 31.713 & 
0.43  & 3 \nl   
C3-V8  & 56.100 & 25.179 & 0.060 & 23.861 & 0.058 & 31.315 & 30.922 & 30.359 & 
0.27  & 3 \nl   
C3-V9  & 48.500 & 24.715 & 0.053 & 23.601 & 0.242 & 30.768 & 30.570 & 30.286 & 
0.63  & 3 \nl   
C3-V10 &  8.800 & 25.785 & 0.098 & 24.721 & 0.439 & 29.792 & 29.421 & 28.892 & 
1.10  & 1 \nl   
C3-V11 & 15.200 & 26.510 & 0.220 & 25.212 & 0.528 & 31.172 & 30.639 & 29.877 & 
1.34  & 4 \nl   
C3-V12 & 54.400 & 25.772 & 0.107 & 24.156 & 0.167 & 31.963 & 31.277 & 30.297 & 
0.49  & 2 \nl   
C3-V13 & 66.600 & 24.988 & 0.061 & 23.631 & 0.053 & 31.421 & 31.021 & 30.450 & 
0.27  & 2 \nl   
C3-V14 & 23.800 & 26.022 & 0.155 & 25.020 & 0.100 & 31.222 & 31.043 & 30.787 & 
0.39  & 4 \nl   
C3-V15 & 13.600 & 26.540 & 0.168 & 25.717 & 0.123 & 31.069 & 30.995 & 30.891 & 
0.44  & 4 \nl   
C3-V16 & 23.800 & 25.825 & 0.112 & 24.822 & 0.154 & 31.004 & 30.822 & 30.562 & 
0.46  & 4 \nl   
C4-V1  & 28.700 & 25.339 & 0.092 & 24.013 & 0.199 & 30.763 & 30.284 & 29.600 & 
0.55  & 4 \nl   
C4-V2  & 37.200 & 25.162 & 0.096 & 23.882 & 0.139 & 30.896 & 30.498 & 29.929 & 
0.42  & 3 \nl   
C4-V3  & 26.900 & 25.605 & 0.191 & 24.600 & 0.152 & 30.951 & 30.785 & 30.547 & 
0.51  & 5 \nl   
C4-V4  & 54.400 & 24.881 & 0.134 & 23.393 & 0.234 & 31.072 & 30.514 & 29.718 & 
0.64  & 4 \nl   
C4-V5  & 14.300 & 26.078 & 0.149 & 24.979 & 0.143 & 30.666 & 30.324 & 29.835 & 
0.46  & 3 \nl   
C4-V6  & 21.600 & 24.451 & 0.096 & 23.339 & 0.253 & 29.534 & 29.232 & 28.802 & 
0.67  & 1 \nl   
C4-V7  & 19.4   & 26.65  &  ---  &   ---  &  ---  &   ---  &   ---  &  ---   & 
---   & 0 \nl
C4-V8  & 17.600 & 24.995 & 0.160 & 23.935 & 0.257 & 29.860 & 29.586 & 29.195 & 
0.70  & 1 \nl   
C4-V9  & 33.000 & 25.198 & 0.126 & 23.841 & 0.233 & 30.789 & 30.298 & 29.596 & 
0.63  & 2 \nl  
C4-V10 & 14.200 & 26.353 & 0.191 & 25.067 & 0.132 & 30.934 & 30.403 & 29.644 & 
0.47  & 5 \nl   
C4-V11 & 19.0   & 25.98  &  ---  &   ---  &  ---  &   ---  &   ---  &  ---   & 
---   & 0 \nl 
C4-V12 & 40.700 & 25.207 & 0.106 & 24.100 & 0.185 & 31.049 & 30.836 & 30.531 & 
0.52  & 2 \nl   
C4-V13 & 37.200 & 25.540 & 0.134 & 23.996 & 0.343 & 31.275 & 30.612 & 29.665 & 
0.88  & 5 \nl   
C4-V14 & 30.200 & 25.933 & 0.119 & 24.353 & 0.087 & 31.418 & 30.692 & 29.655 & 
0.35  & 5 \nl   
C4-V15 & 22.200 & 25.837 & 0.140 & 24.969 & 0.145 & 30.953 & 30.899 & 30.822 & 
0.46  & 3 \nl   
C4-V16 & 32.400 & 25.471 & 0.111 & 24.098 & 0.415 & 31.040 & 30.530 & 29.803 & 
1.04  & 3 \nl   
C4-V17 & 15.600 & 25.523 & 0.125 & 24.799 & 0.338 & 30.216 & 30.260 & 30.321 & 
0.87  & 2 \nl   
C4-V18 & 29.200 & 25.954 & 0.103 & 24.551 & 0.113 & 31.398 & 30.845 & 30.055 & 
0.38  & 3 \nl   
C4-V19 & 26.500 & 26.333 & 0.199 & 24.785 & 0.269 & 31.661 & 30.950 & 29.933 & 
0.74  & 2 \nl   
C4-V20 & 48.500 & 25.310 & 0.091 & 24.010 & 0.130 & 31.363 & 30.979 & 30.430 & 
0.41  & 2 \nl   
C4-V21 & 26.900 & 25.628 & 0.126 & 24.591 & 0.181 & 30.974 & 30.776 & 30.493 & 
0.52  & 4 \nl   
C4-V22 & 40.700 & 24.853 & 0.067 & 23.782 & 0.076 & 30.695 & 30.517 & 30.262 & 
0.30  & 3 \nl   
C4-V23 & 61.900 & 25.120 & 0.052 & 23.696 & 0.074 & 31.524 & 31.054 & 30.382 & 
0.29  & 2 \nl  
C4-V24 & 18.600 & 25.526 & 0.090 & 24.532 & 0.080 & 30.430 & 30.226 & 29.935 & 
0.32  & 2 \nl   
C4-V25 & 11.600 & 26.489 & 0.188 & 25.258 & 0.249 & 30.827 & 30.325 & 29.608 & 
0.70  & 3 \nl   
C4-V26 & 22.200 & 26.231 & 0.149 & 25.033 & 0.157 & 31.347 & 30.963 & 30.414 & 
0.49  & 4 \nl   
\enddata
\end{deluxetable}
%
\begin{deluxetable}{ccccccc}
\tablecaption{DoPHOT and ROMAFOT mean magnitudes in $V$ and $I$ of 66
  Cepheid candidates.\tablenotemark{a}\label{tbl5}}
\scriptsize
\tablewidth{0pt}
\tablehead{
  & \multicolumn{3}{c}{DoPHOT} & \multicolumn{3}{c}{ROMAFOT} \\
  Variable ID & 
  $<\!V\!>$ & $<\!I\!>$ & Quality Index & 
  $<\!V\!>$ & $<\!I\!>$ & Quality Index \\
  {(1)} & {(2)} & {(3)} & {(4)} & 
  {(5)} & {(6)} & {(7)}
}
\startdata
C1-V1   & 25.080  & 23.630  & 3  &  25.17  &  23.69 &  3 \\  
C1-V2   & 26.390  & 25.490  & 3  &  26.49  &  25.58 &  4 \\  
C1-V3   & 25.310  & 25.085  & 3  &  25.44  &  25.34 &  1 \\  
C1-V4   & 26.069  & 25.189  & 5  &  26.38  &  25.42 &  4 \\  
C1-V5   & 25.718  & 24.500  & 5  &  25.81  &  24.67 &  4 \\  
C1-V6   & 25.825  & 25.515  & 3  &  26.05  &  25.63 &  2 \\  
C1-V7   & 25.177  & 23.728  & 3  &  25.23  &  24.14 &  4 \\  
C1-V8   & 26.137  & 25.004  & 4  &  26.05  &  25.18 &  2 \\  
C1-V9   & 26.664  & 25.869  & 1  &  27.19  &  25.99 &  2 \\  
C1-V10  & 25.150  & 24.073  & 3  &  25.23  &  24.08 &  4 \\  
C1-V11  & 26.894  & 25.204  & 5  &  26.76  &  25.06 &  5 \\  
C1-V12  & 25.74   &   ---   & -  &  25.45  &  24.24 &  2 \\     
C1-V13  & 25.541  & 25.434  & 3  &  25.03  &  24.57 &  2 \\  
C1-V14  & 26.704  & 25.807  & 4  &  26.81  &  25.87 &  2 \\  
C1-V15  & 26.116  & 25.349  & 3  &  26.50  &  25.73 &  3 \\  
C1-V16  & 26.307  & 25.310  & 2  &  26.45  &  25.43 &  2 \\  
C1-V17  & 25.990  & 24.893  & 2  &  26.20  &  25.07 &  4 \\  
C1-V18  & 26.586  & 25.817  & 3  &  26.66  &  26.12 &  2 \\  
C2-V1   & 25.390  & 24.258  & 3  &  25.47  &  24.32 &  4 \\  
C2-V2   & 25.921  & 24.954  & 5  &  26.00  &  25.15 &  4 \\  
C2-V3   & 26.276  & 25.216  & 3  &  26.27  &  25.41 &  3 \\  
C2-V4   & 26.88   &   ---   & -  &  27.26  &   ---  &  - \\  
C2-V5   & 26.752  & 26.182  & 4  &  26.96  &  25.94 &  2 \\  
C2-V6   & 26.901  & 25.569  & 5  &  27.03  &  25.81 &  5 \\  
C3-V1   & 26.274  & 25.195  & 5  &  26.45  &  25.58 &  5 \\  
C3-V2   & 25.466  & 24.375  & 5  &  25.65  &  24.56 &  5 \\  
C3-V3   & 26.583  & 26.116  & 1  &  26.77  &  26.27 &  1 \\  
C3-V4   & 25.221  & 23.837  & 2  &  25.32  &  23.76 &  3 \\  
C3-V5   & 25.364  & 24.132  & 1  &  25.57  &  24.09 &  1 \\  
C3-V6   & 25.916  & 24.464  & 2  &  25.97  &  24.72 &  3 \\  
C3-V7   & 24.952  & 24.237  & 3  &  25.01  &  24.51 &  2 \\  
C3-V8   & 25.229  & 23.911  & 3  &  25.30  &  24.02 &  4 \\  
C3-V9   & 24.765  & 23.651  & 3  &  24.80  &  23.80 &  4 \\  
C3-V10  & 25.835  & 24.771  & 1  &  25.78  &  24.91 &  1 \\  
\tableline
\tablebreak
C3-V11  & 26.560  & 25.262  & 4  &  26.50  &  25.51 &  2 \\  
C3-V12  & 25.822  & 24.206  & 2  &  26.06  &  24.28 &  3 \\  
C3-V13  & 25.038  & 23.681  & 2  &  25.16  &  23.80 &  3 \\  
C3-V14  & 26.072  & 25.070  & 4  &  26.17  &  25.15 &  4 \\  
C3-V15  & 26.590  & 25.767  & 4  &  26.77  &  25.82 &  4 \\  
C3-V16  & 25.875  & 24.872  & 4  &  26.03  &  25.16 &  5 \\  
C4-V1   & 25.389  & 24.063  & 4  &  25.34  &  24.11 &  4 \\  
C4-V2   & 25.212  & 23.932  & 3  &  25.03  &  23.87 &  1 \\  
C4-V3   & 25.655  & 24.650  & 5  &  26.17  &  25.62 &  2 \\  
C4-V4   & 24.931  & 23.443  & 4  &   ---   &   ---  &  - \\  
C4-V5   & 26.128  & 25.029  & 3  &  25.97  &  24.64 &  2 \\  
C4-V6   & 24.501  & 23.389  & 1  &  24.43  &  23.20 &  3 \\  
C4-V7   & 26.70   &   ---   & -  &   ---   &   ---  &  - \\
C4-V8   & 25.045  & 23.985  & 1  &  25.43  &   ---  &  - \\  
C4-V9   & 25.248  & 23.891  & 2  &  25.49  &  24.56 &  3 \\  
C4-V10  & 26.403  & 25.117  & 5  &  26.07  &  25.22 &  1 \\  
C4-V11  & 26.03   &   ---   & -  &  25.67  &  23.91 &  2 \\  


C4-V12  & 25.257  & 24.150  & 2  &  25.27  &  24.78  &  4  \\  
C4-V13  & 25.590  & 24.046  & 5  &  25.65  &  25.09  &  5  \\  
C4-V14  & 25.983  & 24.403  & 5  &  26.09  &  24.63  &  5  \\  
C4-V15  & 25.887  & 25.019  & 3  &  26.04  &  24.98  &  3  \\  
C4-V16  & 25.521  & 24.148  & 3  &  25.26  &  23.97  &  3  \\  
C4-V17  & 25.573  & 24.849  & 2  &  25.53  &   ---   &  -  \\  
C4-V18  & 26.004  & 24.601  & 3  &  25.85  &  24.52  &  2  \\  
C4-V19  & 26.383  & 24.835  & 2  &  26.55  &  24.88  &  2  \\  
C4-V20  & 25.360  & 24.060  & 2  &  25.23  &  24.09  &  4  \\  
C4-V21  & 25.678  & 24.641  & 4  &  25.97  &  24.87  &  4  \\  
C4-V22  & 24.903  & 23.832  & 3  &  24.97  &  24.11  &  4  \\  
C4-V23  & 25.170  & 23.746  & 2  &  25.19  &  23.73  &  5 \\  
C4-V24  & 25.576  & 24.582  & 2  &  25.55  &  24.71  &  3  \\  
C4-V25  & 26.539  & 25.308  & 3  &  26.59  &  25.55  &  2  \\  
C4-V26  & 26.281  & 25.083  & 4  &  26.39  &  25.67  &  4  \\  
\enddata
\tablenotetext{a}{All values are on long exposure scale.}
\end{deluxetable}
\begin{table}
\begin{center}
\caption{The average differences between the ROMAFOT and DoPHOT photometry. }
\label{tbl6}
\begin{tabular}{ccccccc}
\noalign{\medskip}
\noalign{\medskip}
\hline\hline
\noalign{\smallskip}
Chip &  \multicolumn{2}{c}{ROMAFOT $-$ DoPHOT}  & Number of &
\multicolumn{2}{c}{ROMAFOT $-$ DoPHOT} & Number of  \\ 
  & $\Delta V$  & $\sigma$ & stars & $\Delta I$ & $\sigma$ & stars\\ 
\noalign{\smallskip}
\hline
\noalign{\smallskip}
1   & $-0.06$ & 0.13 & 106   & $-0.06$ & 0.10 & 143 \\
2   & $-0.03$ & 0.12 &  73   & $+0.01$ & 0.12 &  77 \\           
3   & $+0.04$ & 0.11 & 105   & $+0.13$ & 0.11 & 130 \\
4   & $+0.07$ & 0.19 & 392   & $+0.13$ & 0.20 & 428 \\
\noalign{\smallskip}
\hline
\end{tabular}
\end{center}
\end{table}

\begin{table}
\begin{center}
\caption{The average differences between the ROMAFOT and DoPHOT Cepheid
candidates. }
\label{tbl6a}
\begin{tabular}{ccccccc}
\noalign{\medskip}
\noalign{\medskip}
\hline\hline
\noalign{\smallskip}
Chip &  \multicolumn{2}{c}{ROMAFOT $-$ DoPHOT}  & Number of &
\multicolumn{2}{c}{ROMAFOT $-$ DoPHOT} & Number of  \\ 
  & $\Delta V$  & $\sigma$ & stars & $\Delta I$ & $\sigma$ & stars\\ 
\noalign{\smallskip}
\hline
\noalign{\smallskip}
1   & $+0.08$ & 0.24 &  18   & $+0.10$ & 0.28 &  17 \\
2   & $+0.14$ & 0.13 &   6   & $+0.12$ & 0.23 &   5 \\           
3   & $+0.11$ & 0.09 &  16   & $+0.15$ & 0.12 &  16 \\
4   & $+0.02$ & 0.21 &  24   & $+0.12$ & 0.28 &  19 \\
\noalign{\smallskip}
\hline
\end{tabular}
\end{center}
\end{table}
\begin{table}
\small
\begin{center}
\caption{Distances of NGC 4527 from DoPHOT and ROMAFOT photometry.} 
\label{tab:ROMAFOT_distances}
\begin{tabular}{llccc}
\noalign{\medskip}
\noalign{\medskip}
\hline\hline
\noalign{\smallskip}
 Method/Selection &  n  & $(m-M)_V$ & $(m-M)_I$ & $(m-M)_0$  \\  
\noalign{\smallskip}
\hline
\noalign{\smallskip} 
{\bf DoPHOT\tablenotemark{a}}  & & & &  \\   [5pt]
\multicolumn{5}{c}{Weighted Averages} \\   [5pt]
$20 \leq P \leq 65$; $QI \geq 3$ & 27 &  &  & $30\fm{48} \pm 0\fm{09}$ \\
$20 \leq P \leq 65$; $QI \geq 3$; chip4 & 12 & & & $30\fm{17} \pm 0\fm{13}$ \\
$20 \leq P \leq 65$; $QI \geq 3$; chips1,2,3 & 15 & & & $30\fm{73} \pm 0\fm{12}$ 
\\
$20 \leq P \leq 65$; $QI \geq 3$; chips1,2,3,4($X>400$) & 19 & & & $30\fm{62} 
\pm
 0\fm{10} $ \\
Color Cut (see \S4.4) & 13 & & & $30\fm{69} \pm 0\fm{12}$ \\ 
Combine (.AND.) above 2 lines & 11 & & & $30\fm{63} \pm 0\fm{13}$ \\ [5pt]
\multicolumn{5}{c}{Unweighted Averages} \\  [5pt]
$20 \leq P \leq 65$; $QI \geq 3$ & 27 & $31\fm{20} \pm 0\fm{06} $ & $30\fm{88}  
\pm 0\fm{06} $ & $30\fm{43} \pm 0\fm{10}$ \\
$20 \leq P \leq 65$; $QI \geq 3$; chip4 & 12 & $31\fm{12} \pm 0\fm{07}$ & 
$30\fm{70} \pm 0\fm{06}$  & $30\fm{13} \pm 0\fm{12}$ \\
$20 \leq P \leq 65$; $QI \geq 3$; chips1,2,3 & 15 & $ 31\fm{27} \pm 0\fm{10} $ & 
$ 31\fm{02} \pm 0\fm{08} $ & $30\fm{67} \pm 0\fm{12}$ \\
$20 \leq P \leq 65$; $QI \geq 3$; chips1,2,3,4($X>400$) & 19 &$31\fm{25} \pm 
0\fm{08}$ & $30\fm{98} \pm 0\fm{07}$  & $30\fm{60} \pm 0\fm{10} $ \\
Color Cut (see \S4.4) & 13 &   &   & $30\fm{71} \pm 0\fm{07} $ \\
Combine (.AND.) above 2 lines & 11 & & & $30\fm{66} \pm 0\fm{08}$ \\ [5pt]
{\bf adopted DoPHOT estimate:} &   & {\boldmath $31\fm{26} \pm 0\fm{08}$} &  
{\boldmath $31\fm{00} \pm 0\fm{07}$} & {\boldmath $30\fm{67} \pm 0\fm{12}$} \\            
\hline
\multicolumn{5}{c}{} \\
{\bf ROMAFOT\tablenotemark{a}}    &   &   & &  \\  [5pt]
\multicolumn{5}{c}{Unweighted Averages} \\  [5pt]
\hspace{2cm} $QI \geq$ 3\tablenotemark{b,c} &  33 &  $31\fm{28} \pm 0\fm{05}$ & 
$31\fm{08} \pm 0\fm{05}$ & $30\fm{79} \pm 0\fm{09}$ \\
\hspace{2cm} $QI \geq$ 3; chips1,2,3\tablenotemark{b,c}  &  20 & $31\fm{35} \pm 0\fm{05}$ & 
$31\fm{11} \pm 0\fm{05}$ & $30\fm{77} \pm 0\fm{08}$ \\
\hspace{2cm} $QI \geq$ 3; chip4\tablenotemark{b,c} & 13 & $31\fm{17} \pm 0\fm{10}$ & $31\fm{03} \pm 
0\fm{11}$ & $30\fm{82} \pm 0\fm{20}$ \\ 
$20 \leq P \leq 65$; $QI \geq$ 3; all chips\tablenotemark{b}  & 29 & $31\fm{29} \pm 0\fm{09}$ &
 $31\fm{05} \pm 0\fm{09}$ & $30\fm{71} \pm 0\fm{13} $ \\
$20 \leq P \leq 65$; $QI \geq$ 3; all chips\tablenotemark{b,c} & 26 & $31\fm{27} \pm 0\fm{05}$ &
 $31\fm{09} \pm 0\fm{06}$ & $30\fm{82} \pm 0\fm{11} $ \\
$20 \leq P \leq 65$; $QI \geq$ 3; chips1,2,3,4($X>400$)\tablenotemark{b} & 20 & $31\fm{42} \pm 0\fm{09}$ & 
$31\fm{16} \pm 0\fm{06}$ & $30\fm{78} \pm 0\fm{09}$ \\
$20 \leq P \leq 65$; $QI \geq$ 3; Color Cut (see \S4.4)\tablenotemark{b} & 16 & $31\fm{26} \pm 0\fm{06}$ & $31\fm{13} \pm 
0\fm{06}$ & $30\fm{93} \pm 0\fm{08}$ \\
$20 \leq P \leq 65$;  $QI \geq$ 3; AC with N4527\tablenotemark{c} &  26 & $31\fm{09} \pm 0\fm{05}$ & $30\fm{95} \pm 
0\fm{06}$ & $30\fm{75} \pm 0\fm{11}$ \\
$20 \leq P \leq 65$; $QI \geq$ 3; AC with Pal4, G319\tablenotemark{c} &  26 & $31\fm{36} \pm 
0\fm{05}$ & $31\fm{14} \pm 0\fm{06}$ & $30\fm{83} \pm 0\fm{11}$ \\
$20 \leq P \leq 65$; $QI \geq$ 3; AC with Pal4\tablenotemark{c} &  26 & $31\fm{33} \pm 
0\fm{05}$ & $31\fm{15} \pm 0\fm{06}$ & $30\fm{88} \pm 0\fm{11}$ \\
$20 \leq P \leq 65$; $QI \geq$ 3; AC with G319\tablenotemark{c} &  26 & $31\fm{39} \pm 
0\fm{05}$ & $31\fm{14} \pm 0\fm{06}$ & $30\fm{78} \pm 0\fm{11}$ \\
\bf adopted ROMAFOT estimate: & {\bf 26} & {\boldmath $31\fm{27} \pm 0\fm{05}$} & 
{\boldmath $31\fm{09} \pm 0\fm{06}$} & {\boldmath $30\fm{82} \pm 0\fm{11}$} \\
\hline 
\noalign{\smallskip}
\bf{final adopted}  &   & {\boldmath $31\fm{27} \pm 0\fm{08}$} & {\boldmath 
$31\fm{05} \pm 0\fm{07}$} & {\boldmath $30\fm{74} \pm 0\fm{12}$} \\ \hline 
\noalign{\smallskip}
\hline
\noalign{\smallskip}
\tablenotetext{a}{All values are on long exposure scale.}
\tablenotetext{b}{The mean AC from NGC 4527, Pal4 and G319 is used.} 
\tablenotetext{c}{The Cepheid candidates C1-V11, C2-V6, C3-V12, C4-V6 are excluded.}
\end{tabular}
\end{center}
\end{table}

\begin{deluxetable}{lclccccc}
\tablecaption{Photometric Parameters of SN\,1991T-like SNe\,Ia.
\label{tbl:SNe}}
\small
\tablewidth{0pt}
\tablehead{ 
  \colhead{SN} & \colhead{log $v$} & \colhead{$\Delta m_{15}$} &
  \colhead{$M_{\rm B}^0$} & \colhead{$M_{\rm V}^0$} &
  \colhead{$M_{\rm I}^0$} & \colhead{$(B-V)^0$} & \colhead{$(V-I)^0$}} 
\startdata
1991T\tablenotemark{a} & --- & 0.95 
     & $-19.86$ & $-19.85$ & $-19.45$ & $-0.01$ & $-0.40$ \\
1995ac\tablenotemark{b} & 4.166 & 0.96\tablenotemark{d} 
     & $-20.04$ & $-19.98$ & $-19.74$ & $-0.06$ & $-0.24$ \\
1995bd\tablenotemark{c} & 3.681 & 0.89\tablenotemark{d} 
     & $-19.39$ & $-19.68$ & --- & $+0.29$ & --- \\
 & & 
     & $(-20.62)$ & $(-20.61)$ & --- & $(-0.01)$ & --- \\
1997br\tablenotemark{e} & 3.342 & 1.00 
     & $-19.25$ & $-19.46$ & $-19.54$ & $+0.21$ & $+0.08$ \\
 & & 
     & $(-20.36)$ & $(-20.28)$ & $(-20.12)$ & $(-0.08)$ & $(-0.16)$ \\
\enddata
\tablenotetext{a}{Luminosities repeated from equation~(\ref{eq:13}).}
\tablenotetext{b}{Apparent magnitudes from \cite{rie99}. Corrected for
  Galactic absorption (Schlegel et~al.\ 1998). $H_0=60$ adopted.}
\tablenotetext{c}{Apparent magnitudes from \cite{rie99}. First entry:
.  corrected only for Galactic absorption (Schlegel et~al.\
.  1998). Second entry: Assuming normal color $(B-V)= -0.01$ and
  correcting for additional internal absorption. $H_0=60$ adopted.}
\tablenotetext{d}{Mean of \cite{rie99} and Phillips et~al.\ (1999).}
\tablenotetext{e}{Photometric parameters from Li et~al.\ (1999). First
  entry: corrected for Galactic absorption (Schlegel et~al.\
  1998). Second entry: corrected for total absorption assuming
  $E(B-V)=0.40$ (cf.\ text). $H_0=60$ adopted.}
\end{deluxetable}




\begin{thebibliography}{}
%
\bibitem[Baade (1938)]{baa38} Baade, W. 1938, \apj, 88, 285
%
\bibitem[Baade \& Zwicky (1938)]{baaz38} Baade, W., \& Zwicky, F. 1938, \apj, 
88, 411
%
\bibitem[Barbon et al.\ (1973)]{bar73} Barbon, R., Ciatti, F., \&  
Rosino, L. 1973, \aap, 25, 241
%
%
\bibitem[Bertola (1964)]{ber64} Bertola, F 1964, Ann d'Ap, 27, 319 
%
\bibitem[Bertola et al.\ (1965)]{ber65} Bertola, F., Mammano, A.S., \&
Perinotto, M. 1965, Contr. Asiago Obs, 174.
%
\bibitem[Bertola \& Sussi (1965)]{bersu65} Bertola, F., \& Sussi, M.G. 1965,
Contr. Oss. Astron. Univ. Padua, No. 176
%
\bibitem[Binggeli, Popescu, \& Tammann 1993]{bipota93}
  Binggeli, B., Popescu, C.C., \& Tammann, G.A. 1993, A\&AS, 98, 275 
%
\bibitem[Branch (1981)]{bra81} Branch, D. 1981, \apj, 248, 1076
%
\bibitem[Branch (1982)]{bra82} Branch, D. 1982, \apj, 258, 35
%
\bibitem[Branch 1986]{bra86} Branch, D. 1986, \apj, 300, L51
%
\bibitem[Branch, Fisher, \& Nugent 1993]{bra93} Branch, D., Fisher A.,
\& Nugent, P. 1993, \aj, 106, 2383
%
\bibitem[Branch, Nugent, \& Fisher 1997]{bra97} Branch, D., Nugent,
  P., \& Fisher A. 1997, in Thermonuclear Supernovae,
  eds. P. Ruiz-Lapuente, R. Canal, \& J. Isern (Dordrecht: Kluwer),
  p.~715 
%
\bibitem[Branch, \& Tammann 1992]{bra92} Branch, D., \& Tammann, G.A. 
1992, \araa, 30, 359
%
\bibitem[Buonanno et~al.\ 1983]{buo83} 
  Buonanno, R., Buscema, G., Corsi, C.E., Ferraro, I., \& Iannicola,
  G. 1983, \aap, 126, 278
%
\bibitem[Caldwell \& Coulson (1985)]{cal85} Caldwell, J.A.R., \& 
Coulson, I.M. 1985, \mnras, 212, 879
%
\bibitem[Dean, Warren, \& Cousins (1978)]{dea78} Dean, J.F., Warren, 
P.R., \& Cousins, A.W.J. 1978, \mnras, 183, 569
%
\bibitem[de Vaucouleurs (1975)]{dev75} de Vaucouleurs, G. 1975 in 
Galaxies and the Universe, eds. A. Sandage, M. Sandage, \& J. Kristian, 
Vol 9 of Stars and Stellar Systems (Chicago: University of Chicago 
Press), Chapter 14, p 557 
%
\bibitem[Elias et al.\ (1985)]{eli85} Elias, J.H., Mathews, K., Neugebauer, G.,
\& Persson, S.E. 1985, \apj, 296, 378.
%
\bibitem[Federspiel, Tammann, \& Sandage 1998]{fed98} Federspiel, M., 
Tammann, G.A., \& Sandage, A. 1998, \apj, 495, 115 
%
\bibitem[Ferrarese et al.\ (2000)]{ferr00} Ferrarese, L., et al. 2000, 
\pasp, 112, 177
%
\bibitem[Fernie (1990)]{fern90} Fernie, J.D. 1990, \apj, 354, 295
%
\bibitem[Filippenko et al.\ (1992)]{fil92} Filippenko, A.V., et al. 1992, 
\apj, 385, L15
%
\bibitem[Fisher et al.\ (1999)]{fish99} Fisher, A., Branch, D., Hatano, K. 
\& Baron, E. 1999, \mnras, 304, 67
%
\bibitem[Garnavich et~al.\ 1996]{gar96} Garnavich, P.M., Riess, A.G.,
  Kirshner, R.P., Challis, P., \& Wagner, R.M. 1996, BAAS, 28, 1331  
%
\bibitem[Garnavich et~al.\ 1997]{gar97} Garnavich, P., Jha, S.,
  Kirshner, R., \& Challis, P. 1997, IAU~Circ. 6699  
%
\bibitem[Garnavich, Jha, \& Kirshner 1998]{gar98} Garnavich, P., Jha, S.,
  \& Kirshner, R. 1998, IAU~Circ. 6857  
%
\bibitem[Gibson et~al.\ 2000]{gib00} Gibson, B.K., et~al.\ 2000, \apj,
  529, 723
%
\bibitem[Hamuy et al.\ 1996]{ham96a} Hamuy, M., Phillips, M.M.,
Schommer, R.A., Suntzeff, N.B., Maza, J., \& Aviles, R. 1996, \aj,
112, 2391
%
\bibitem[Harkness et al.\ 1987]{har87} Harkness, R.P., et al. 1987, 
\apj, 317, 355
%
\bibitem[Harkness \& Wheeler (1990)]{harwh90} Harkness, R.P. \& Wheeler, J.C.
 1990, in Supernovae, ed. A.G. Petschek (Springer)
%
%
\bibitem[H\"oflich 1995]{hof95} H\"oflich, P.A. 1995, \apj, 443, 89
%
\bibitem[Holtzman et al.\ 1995a]{hol95a} Holtzman, J.A., et al.\ 1995a, 
\pasp, 107, 156
%
\bibitem[Holtzman et al.\ (1995b)]{hol95b} Holtzman, J.A., et al.\ 1995b, 
\pasp, 107, 1065
%
\bibitem[Jha et~al.\ 1998]{jha98}
   Jha, S., Garnavich, P., Challis, P., \& Kirshner, R. 1998,
   IAU~Circ. 7054
%
\bibitem[Kowal 1968]{kow68} Kowal, C.T. 1968, \aj, 73, 1021
%
\bibitem[Labhardt, Sandage, \& Tammann (1997)]{lab97} Labhardt, L., 
Sandage, A., \& Tammann, G.A. 1997, \aap, 322, 751 
%
\bibitem[Lafler-Kinman algorithm (1965)]{laf65} Lafler, J. \& Kinman, 
T.D. 1965, \apjs, 11, 216
%
\bibitem[Li et~al.\ (1999)]{li99}
   Li, W.D., et~al.\ 1999, \aj, 117, 2709
%
\bibitem[Lira et~al.\ (1998)]{lira98} Lira, P., et~al. 1998, \aj, 115, 234
%
\bibitem[Livio et al.\ 2000]{liv00} Livio, M., Panagia, N. \& Sahu, K. 2000,
eds of {\it The Greatest Explosions since the Big Bang: Supernovae and 
Gamma-Ray Bursts} (Cambridge University Press), in press
%
\bibitem[Madore \& Freedman 1991]{mad91} Madore, B.F., \& Freedman,
W.L.  1991, \pasp, 103, 933
%
%
\bibitem[Mazzali, Danziger, \& Turatto 1995]{maz95}
   Mazzali, P.A., Danziger, I.J., \& Turatto, M. 1995, \aap, 297, 509
%
\bibitem[Minkowski (1939)]{min39} Minkowski, R. 1939, \apj, 89, 156
%
\bibitem[Minkowski 1941]{min41} Minkowski, R. 1941, \pasp, 53, 224
%
\bibitem[Minkowski 1964]{min64} Minkowski, R. 1964, \araa, 2, 247
%
\bibitem[Mochejska et al.\ (2000)]{Moch99} Mochejska, B.J., Macri, L.M., 
Sasselov, D.D., \& Stanek, K.Z. 2000, AJ, submitted, astro-ph/9908293
%
\bibitem[Nugent et~al.\ 1995]{nug95}
   Nugent, P., Phillips, M., Baron, E., Branch, D., \& Hauschildt,
   P. 1995, \apj, 455, L147
%
\bibitem[Panagia 1985]{pan85} Panagia, N. 1985, in Supernovae as 
Distance Indicators; Lecture Notes in Physics No. 224, ed. N. Bartel (New York: 
Springer), p14
%
\bibitem[Parodi et al.\ 2000]{par00} Parodi, B.R., Saha, A., Sandage,
A., \& Tammann, G.A. 2000, \apj, in press
%
%
\bibitem[Phillips (1993)]{phi93} Phillips, M.M. 1993, \apjl, 413, L105
%
%
\bibitem[Phillips et al.\ (1992)]{phi92} Phillips, M.M., et al.1992, 
\aj, 103, 1632 
%
\bibitem[Phillips et al.\ 1999]{phi99} Phillips, M.M., et al.\ 1999, 
\aj, 118, 1766
%
\bibitem[Porter \& Filippenko (1987)]{porfi87} Porter, A.C. \& Filippenko, A.C.
 1987, \aj, 93, 1372
%
\bibitem[Pskovskii (1967)]{psk67} Pskovskii, Yu. P. 1967, SvA, 11, 63
%
\bibitem[Pskovskii (1971)]{psk71} Pskovskii, Yu. P. 1971, SvA, 14, 798
 %
\bibitem[Pskovskii (1984)]{psk84} Pskovskii, Yu. P. 1984, SvA, 28, 658
%
\bibitem[Qiao et~al.\ 1997]{qiao97}
   Qiao, Q.Y., Wu, H., Wei, J.Y., \& Li, W.D. 1997, IAU~Circ. 6623
%
%
\bibitem[Riess et~al.\ (1999)]{rie99} Riess, A.G., et~al.\ 1999, \aj, 117, 707
%
\bibitem[Rizzi et~al.\ 1999]{riz99}
   Rizzi, L., Patat, F., Benetti, S., Cappellaro, E., \& Turatto,
   M. 1999, IAU~Circ. 7215
%
\bibitem[Ruiz-Lapuente et al.\ (1992)]{rla92} Ruiz-Lapuente, P., et al.\ 1992, 
\apj, 387, L387
%
\bibitem[Saha \& Hoessel (1990)]{saha90} Saha, A., \& Hoessel, J.G. 
1990, \aj, 99, 97
%
\bibitem[Saha et al.\ 1996a]{saha96a} Saha, A., Sandage, A., Labhardt,
L., Tammann, G.A., Macchetto, F.D., \& Panagia, N.  1996a, \apj, 466,
55 (Paper~V; NGC~4536; SN~1981B)
%
\bibitem[Saha et al.\ 1996b]{saha96b} Saha, A., Sandage, A., Labhardt,
L., Tammann, G.A., Macchetto, F.D., \& Panagia, N.  1996b, \apjs, 107,
693 (Paper~VI; NGC~4496A; SN~1960F)
%
\bibitem[Saha et al.\ 1997]{saha97} Saha, A., Sandage, A., Labhardt,
L., Tammann, G.A., Macchetto, F.D., \& Panagia, N. 1997, \apj, 486, 1
(Paper~VIII; NGC~4639; SN~1990N)
%
\bibitem[Saha et al.\ 1999]{saha99} Saha, A., Sandage, A., Tammann, G.A.,
Labhardt, L., Macchetto, F.D., \& Panagia, N. 1999, \apj, 522, 802
(Paper~IX; NGC~3627; SN~1989B)
%
\bibitem[Saha et al.\ 2000]{saha00} Saha, A., Labhardt, L., Prosser, C. 
2000, \pasp, 112, 163.
%
\bibitem[Sandage 1988]{san88} Sandage, A. 1988, \pasp, 100, 935
%
\bibitem[Sandage, Bell, \& Tripicco (1999)]{sanetal99} Sandage, A., 
Bell, R.A., \& Tripicco, M.J. 1999, \apj, 522, 250
%
\bibitem[Sandage et al.\ 1996]{sanetal96} Sandage, A., Saha, A.,
Tammann, G.A., Labhardt, L., Panagia, N., \& Macchetto, F.D.  1996,
\apjl, 460, L15 (Paper~VII; NGC~4639; SN~1990N)
%
\bibitem[Sandage \& Tammann 1968]{sata68} Sandage, A., \& Tammann,
G.A. 1968, \apj, 151, 531
%
\bibitem[Sandage \& Tammann (1982)]{sata82} Sandage, A., \& Tammann,
G.A. 1982, \apj, 256, 339 (Steps VII)
%
\bibitem[Sandage \& Tammann 1987]{sata87} Sandage, A., \& Tammann,
G.A. 1987, A Revised Shapley-Ames Catalog of Bright Galaxies, 2nd ed.,
(Washington: Carnegie Institution), p.~47 
%
\bibitem[Sandage \& Tammann (1993)]{sata93} Sandage, A., \& Tammann,
G.A. 1993, \apj, 415, 1
%
\bibitem[Sandage \& Tammann (1997)]{sata97} Sandage, A., \& Tammann,
G.A. 1997, in Critical Dialogues in Cosmology, ed. N. Turok (Singapore; World
Scientific), p130
%
\bibitem[Sandage, Tammann, \& Saha 2000]{satasa00} Sandage, A., Tammann,
G.A., \& Saha, A. 2000, in {\it The Greatest Explosions since the Big Bang: 
Supernovae and Gamma-Ray Bursts}, eds. M.~Livio, N.~Panagia \& K.~Sahu
(Cambridge University Press), in press.
\bibitem[Schechter et al.\ 1993]{schec93} Schechter, P.L., Mateo, 
M.L., \& Saha, A. 1993, \pasp, 105, 1342
%
\bibitem[Scheffler 1982]{schef82}  Scheffler, H. 1982, in 
Landolt-B\"{o}rnstein, Astronomy \& Astrophysics, vol. 2c, 
eds. K. Schaifers \& H.H. Voigt (Berlin: Springer), p46
%
\bibitem[Schlegel et~al. (1998)]{schle98} Schlegel, D.J., Finkbeiner,
  D.P., \& Davis, M.  1998, \apj, 500, 525
%
\bibitem[Sparks et~al. (1999)]{sparks99} Sparks, W.B., Macchetto, F., 
Panagia, N., Boffi, F.R., Branch, D., Hazen, M.L., \& della Valle, M. 1999,
\apj, 523, 585
%
\bibitem[Stetson (1995)]{ste95} Stetson, P.B. 1995, private communication  
%
\bibitem[Stanek et al.\ (2000)]{Staudal99} Stanek, K.Z., \& Udalski, A.
2000,  ApJL, submitted,  astro-ph/9909346
%
\bibitem[Tammann \& Sandage 1995]{tasa95} Tammann, G.A., \& Sandage, A. 
    1995, \apj, 452, 16
%
\bibitem[Tripp 1998]{tri98} Tripp, R. 1998, \aap, 331, 815
%
\bibitem[Tripp \& Branch 1999]{tribra99} Tripp, R., \& Branch, D. 1999, \apj, 
525, 209
%
\bibitem[Tully (1988)]{tul88} Tully, R.B. 1988, {\it Nearby Galaxy Catalog}
(Cambridge University Press).
%
\bibitem[Tully \& Shaya 1984]{tul84} Tully, R.B., \& Shaya, E.J. 1984,
  \apj, 281,31.
%
\bibitem[Uomoto \& Kirshner 1985]{uokir85} Uomoto, A. \& Kirshner, R.P. 1985,
\aap, 148, L7
%
\bibitem[Wheeler \& Levreault 1985]{wheelev85} Wheeler, J.C., \& Levreault, 
R. 1985, \apj, 313, L69
%
\end{thebibliography}
\end{document}